\documentclass[aps,prd,onecolumn,superscriptaddress,nofootinbib,floatfix,11pt]{revtex4-2}

\usepackage{graphicx} 
\usepackage{xcolor}
\usepackage{subfigure}
\usepackage{float}
\usepackage{amsmath}
\usepackage{amssymb}
\usepackage{slashed}
\usepackage{braket}
\usepackage{breqn}
\usepackage{verbatim}
\usepackage[utf8]{inputenc} 
\allowdisplaybreaks
\usepackage[
colorlinks=true,
linkcolor=blue,
breaklinks=true,
urlcolor=blue,
citecolor=blue]{hyperref}

\usepackage{appendix}
\usepackage{orcidlink}

\newcommand{\pku}{\affiliation{School of Physics, Peking University, Beijing 100871, China}}

\newcommand{\itp}{\affiliation{Institute of Theoretical Physics, Chinese Academy of Sciences, Beijing 100190, China}}

\newcommand{\ucas}{\affiliation{School of Physical Sciences, University of Chinese Academy of Sciences, Beijing 100049, China}}

\newcommand{\scnt}{\affiliation{Southern Center for Nuclear-Science Theory (SCNT), Institute of Modern Physics,\\ 
Chinese Academy of Sciences, Huizhou 516000, China}}

\begin{document}
\title{Decays of $\Upsilon(10860)$ and $\Upsilon(10753)$ into $\omega\chi_{bJ}$}

\begin{abstract}
    In this paper, we model $\Upsilon(10753)$ as a $4S$-$3D$ bottomonium mixture and $\Upsilon(10860)$ as a $5S$-$4D$ mixture, and predict the properties of a new bottomonium state, $\Upsilon(10950)$, as the mixing partner of $\Upsilon(10860)$. The mixing angles are derived from dielectron decay widths and mass shifts of these bottomonia. 
    We consider open-bottom meson loops in the decays of the $D$-wave bottomonium components, based on a nonrelativistic effective field theory power counting.
    We show that the $S$-$D$ mixing scheme is consistent with the experimental data of the decays of $\Upsilon(10860)$ into $\omega\chi_{bJ}$ ($J=0,1,2$) and $\Upsilon(10753)$ into $\omega\chi_{bJ}$ ($J=1,2$).
    Predictions for the dielectron widths and partial decay widths into $\chi_{bJ}\omega$ for $\Upsilon(10753)$ and $\Upsilon(10950)$ are presented.
\end{abstract}

\author{Zheng-Li Luo\orcidlink{0009-0001-7326-634X}}\email{luozhengli@stu.pku.edu.cn}
\pku \ucas 

\author{Yi-Lin Song\orcidlink{0009-0005-9302-1349}}\email{songyilin@itp.ac.cn}
\itp \ucas 

\author{Feng-Kun Guo\orcidlink{0000-0002-2919-2064}}\email{fkguo@itp.ac.cn}
\itp \ucas \scnt


\maketitle

\section{Introduction}

The $J^{PC} = 1^{--}$ bottomonia family, known as $\Upsilon$, was initially discovered at Fermilab in 1977 with the identification of its $1S$, $2S$, and $3S$ states~\cite{E288:1977xhf,E288:1977efs}. About thirty years later, a $D$-wave excited state, $\Upsilon_2(1D)$, was observed~\cite{CLEO:2004npj,BaBar:2010tqb}. 
The $\Upsilon$ states below the open-bottom  threshold align well with the predictions of the Godfrey-Isgur quark model~\cite{Godfrey:1985xj}; however, above this threshold, the $\Upsilon$ spectrum exhibits complexities~\cite{Brambilla:2010cs, Drutskoy:2012gt}. 

In 2019, Belle reported a bottomonium-like resonant structure, $\Upsilon(10753)$, in the invariant mass distribution of $\Upsilon(nS)\pi^+\pi^-$ $(n = 1,2,3)$~\cite{Belle:2019cbt}.
As mentioned by the Particle Data Group (PDG), it is a candidate for a conventional $3D$ state or an exotic structure~\cite{ParticleDataGroup:2024cfk}.
However, its mass exceeds sizably the predicted mass range for the $3D$ state (e.g., 10653~MeV~\cite{Segovia:2016xqb}, 10675~MeV~\cite{Wang:2018rjg}, and 10698~MeV~\cite{Godfrey:1985xj,Godfrey:2015dia}).
Consequently, a $4S$-$3D$ mixture model has been proposed, with predictions for the decay widths $\Upsilon(10753)\to \pi^+\pi^-\Upsilon(nS)$ and $\Upsilon(10753)\to\omega\chi_{bJ}(1P)$~\cite{Li:2021jjt,Bai:2022cfz}.
Mixing between two energy levels always causes them to repel each other. 
Thus, in such a mixing scenario, the $4S$ and $3D$ states, which were predicted to have masses of 10635~MeV and 10698~MeV, respectively, in the Godfrey-Isgur quark model~\cite{Godfrey:1985xj,Godfrey:2015dia}, mix to produce the $\Upsilon(10580)$ and $\Upsilon(10753)$.

As for the exotic interpretation, the $\Upsilon(10753)$ has been described as a hybrid~\cite{TarrusCastella:2021pld} or a tetraquark state~\cite{Wang:2019veq,Ali:2019okl}.
Furthermore, there is another member of the $\Upsilon$ family, $\Upsilon(10860)$, located approximately 100~MeV above. This state also requires further investigation. 
Despite $\Upsilon(10860)$ being within the $5S$ bottomonium mass range predicted in quark models~\cite{Godfrey:1985xj,Godfrey:2015dia,Segovia:2016xqb,Wang:2018rjg,Chen:2025zbu}, the heavy quark spin symmetry (HQSS) breaking in its decays into open-bottom mesons is unexpectedly large~\cite{Drutskoy:2012gt, Mehen:2013mva,Voloshin:2013ez} in the $5S$ description. 
This has led to the consideration of a small $D$-wave component in its wave function~\cite{Guo:2014qra}. 
An upper limit for the $5S$-$4D$ mixing angle of $\Upsilon(10860)$ was estimated to be $27^\circ$ based on the experimental dielectron width~\cite{Badalian:2009bu}. 

In Ref.~\cite{Belle-II:2022xdi}, the Belle~II experiment reported the first observation of $\omega\chi_{bJ}\,(J=1,2)$ at a center-of-mass (c.m.) energy of 10.745~GeV.\footnote{No evidence was found for the radiative processes $e^+e^-\to \gamma\chi_{bJ}\,(J=0,1,2)$ at the same energy~\cite{Belle-II:2025iil}.} The energy dependence of the Born cross section for $e^+e^-\to \omega\chi_{bJ}\,(J=1,2)$ was fitted using a coherent sum of two-body phase space and a Breit-Wigner function, providing insights into the internal structure of the $\Upsilon(10753)$. 
Two solutions were found for the product of the $\Upsilon(10753)$ dielectron width and the $\omega \chi_{bJ}\,(J=1,2)$ branching fractions~\cite{Belle-II:2022xdi}:
\begin{align}
    \Gamma_{ee}\mathcal{B}(\Upsilon(10753)\to \omega\chi_{b1}) &= (0.63\pm 0.39_{\text{stat.}}\pm 0.20_{\text{sys.}})\,\text{eV}\notag,\\
    \Gamma_{ee}\mathcal{B}(\Upsilon(10753)\to \omega\chi_{b2}) &= (0.53\pm 0.46_{\text{stat.}}\pm 0.15_{\text{sys.}}) \,\text{eV},\label{10753r1}
\end{align}
and 
\begin{align}
    \Gamma_{ee}\mathcal{B}(\Upsilon(10753)\to \omega\chi_{b1}) &= (2.01\pm 0.38_{\text{stat.}}\pm 0.76_{\text{sys.}})\,\text{eV}\notag,\\
    \Gamma_{ee}\mathcal{B}(\Upsilon(10753)\to \omega\chi_{b2}) &= (1.32\pm 0.44_{\text{stat.}}\pm 0.55_{\text{sys.}}) \,\text{eV}.\label{10753r2}
\end{align}
The former corresponds to a constructive interference between the Breit-Wigner term for the $\Upsilon(10753)$ and a phase-space background term, while the latter corresponds to a destructive interference.
Here, $\Gamma_{ee}$ represents the dielectron width of $\Upsilon(10753)$.

The branching fractions for the $\Upsilon(10860)$ decays into $\omega\chi_{bJ}\,(J=0,1,2)$ were measured by Belle a decade ago~\cite{Belle:2014sys}, yielding the following results:
\begin{align}
    \mathcal{B}(\Upsilon(10860)\to \omega\chi_{b0}) &< 3.9\times 10^{-3},\notag\\
    \mathcal{B}(\Upsilon(10860)\to \omega\chi_{b1} )&= (1.57\pm 0.22_{\text{stat.}}\pm 0.21_{\text{sys.}})\times 10^{-3},\notag\\
    \mathcal{B}(\Upsilon(10860)\to \omega\chi_{b2} )&= (0.60\pm 0.23_{\text{stat.}}\pm 0.15_{\text{sys.}})\times 10^{-3}\label{10860r}.
\end{align}
    
Define the ratios of the branching fractions as
\begin{equation}
    R_{ij} = \frac{\Gamma(\Upsilon(10753)\to \omega\chi_{bi})}{\Gamma(\Upsilon(10753)\to \omega\chi_{bj})},\qquad R_{ij}' = \frac{\Gamma(\Upsilon(10860)\to \omega\chi_{bi})}{\Gamma(\Upsilon(10860)\to \omega\chi_{bj})},\quad i,j=0,1,2.
\end{equation}
The obtained values are $R_{12}=1.19^{+1.36}_{-1.19}$ (constructive) or $1.52\pm1.04$ (destructive), while $R_{12}' = 2.62\pm 1.30$. 
If the $\Upsilon(10860)$ is a pure $S$-wave bottomonium state, $R_{12}'$ would be 0.64 following from HQSS~\cite{Guo:2014qra}.
In the following section, we will demonstrate that the measured $R_{12}$ is also inconsistent with the $D$-wave scenario for the $\Upsilon(10753)$.

In this work, we will consider $4S$-$3D$ and $5S$-$4D$ mixing scenarios for $\Upsilon(10753)$ and $\Upsilon(10860)$, respectively, with mixing angles determined from dielectron widths and masses. We will then investigate the decays of $\Upsilon(10753)$ and $\Upsilon(10860)$ into $\omega\chi_{bJ}$ ($J=0,1,2$), considering HQSS and open-bottom loop effects.

\section{Formalism}\label{sec2}

We start by summarizing the theoretical framework outlined in Ref.~\cite{Guo:2014qra}.
The framework is based on nonrelativistic effective field theory (NREFT) with HQSS, and the power counting scheme is based on the typical velocity $v$ of the heavy mesons, as elaborated in Refs.~\cite{Guo:2010ak, Guo:2012tg, Guo:2017jvc}.

\begin{figure}[tb]
    \centering
    \includegraphics[width=7cm]{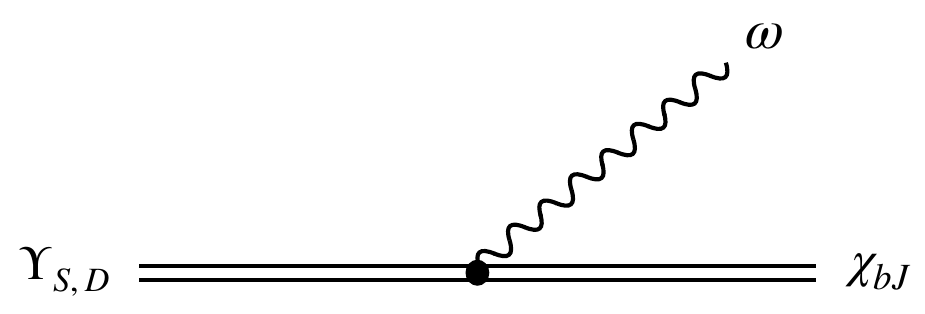}
    \caption{Short-range contribution for the transitions $\Upsilon_{S,\,D}\to \omega \chi_{bJ}$.}
    \label{fig:tree_diagram}
\end{figure}

Since the $\chi_{bJ}$ and $\omega$ can couple to both the $S$-wave and $D$-wave bottomonia in $S$ waves, the leading order effective Lagrangian for the couplings in Fig.~\ref{fig:tree_diagram} is given by
\begin{equation}
    \mathcal{L}_{\chi\omega}=\frac{c_S}{2}\langle \chi^{i\dagger}\Upsilon\rangle\omega^i+\frac{c_D}{2}\langle \chi^{i\dagger}\Upsilon^{ij}\rangle\omega^j + \text{h.c.},
    \label{treeEQ}
\end{equation}
where $ \Upsilon=\Upsilon_S^i\sigma^i$,
$\chi^i=\sigma^j(\frac{1}{\sqrt{3}}\delta^{ij}\chi_0-\frac{1}{\sqrt{2}}\epsilon^{ijk}\chi_1^k-\chi_2^{ij})$ and 
$\Upsilon^{ij}=\frac{3}{2\sqrt{15}}(\Upsilon_D^i\sigma^j+\Upsilon_D^j\sigma^i)-\frac{1}{\sqrt{15}}\delta^{ij}\Upsilon_D^k\sigma^k$ represent fields for the $S$-wave, $P$-wave and $D$-wave bottomonia, respectively~\cite{Guo:2014qra,Casalbuoni:1992yd,Hu:2005gf,Guo:2010ak,margaryanUsingDecayPs41602013}, $\sigma^i$ $(i=1,2,3)$ are the Pauli matrices, and $\langle\cdot\rangle$ denotes the trace over the spinor space. 
The fields for the $S$-wave and $D$-wave bottomonia are denoted by $\Upsilon_S$ and $\Upsilon_D$, and those for $\chi_{bJ}$ and $\omega$ are denoted by $\chi_J$ and $\omega$, respectively. 

With the above Lagrangian, one can easily derive the ratios of decay widths of an $S$-wave or $D$-wave $\Upsilon$ into $\chi_{bJ}\omega$ if we neglect the phase space differences, denoted as $\Gamma_J^S$ and $\Gamma_J^D$, respectively~\cite{Guo:2014qra}. The results for the $S$-wave $\Upsilon$ are given by
\begin{equation}
    \Gamma_0^S:\Gamma_1^S:\Gamma_2^S = 1:3:5,
    \label{1:3:5}
\end{equation}
and the results for the $D$-wave $\Upsilon$ are given by
\begin{equation}
    \Gamma_0^D:\Gamma_1^D:\Gamma_2^D = 20:15:1.
    \label{20:15:1}
\end{equation}
Considering the phase space effects does not change the pattern of the decay width ratios. Comparing with the $R_{12}$ and $R_{12}'$ values extracted above from the experimental results in Eqs.~(\ref{10753r1}), (\ref{10753r2}) and \eqref{10860r}, one concludes that none of the $\Upsilon(10753)$ and $\Upsilon(10860)$ should be a pure $D$- or $S$-wave bottomonium, although the dielectron width of the latter is consistent with the $5S$ bottomonium scenario within errors~\cite{Badalian:2009bu}.

\begin{figure}[tb]
    \centering
    \includegraphics[width=10cm]{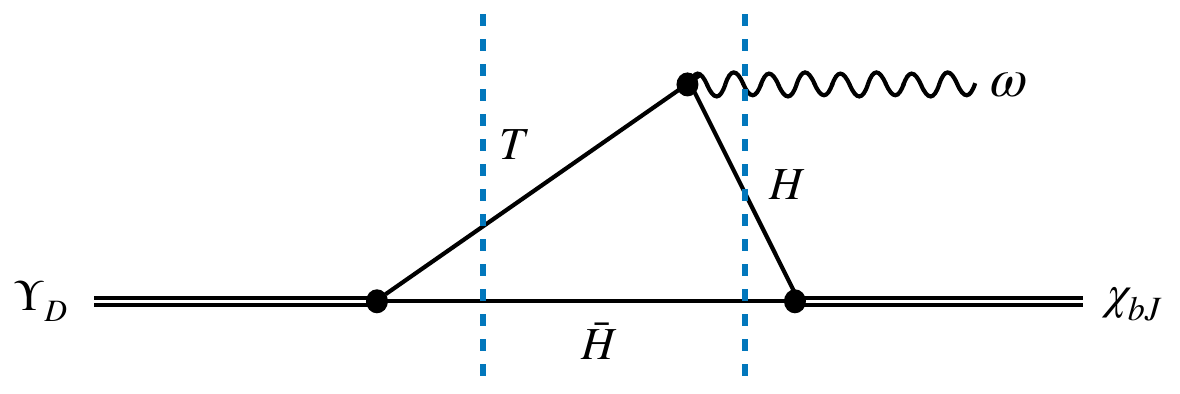}
    \caption{Dominant coupled-channel effects for $\Upsilon_D$ decaying into $\omega \chi_{bJ}$. The dashed lines represent two unitary cuts. The charge conjugated diagram is not shown.}
    \label{fig:triangle_diagram}
\end{figure}

\begin{table}[tb]
    \centering
    \caption{Dominant heavy meson loops for $\Upsilon_D$ decaying into $\omega \chi_{bJ}$. The charge conjugated contributions are also considered but not listed here.}
    \begin{tabular}{lccccc}
    \hline
    Processes & {$\Upsilon_D\rightarrow\chi_{b0}\omega$} & {$\Upsilon_D\rightarrow\chi_{b1}\omega$} & {$\Upsilon_D\rightarrow\chi_{b2}\omega$} \\
    \hline
    $T\bar HH$ & $[B_1\bar BB],[B_1\bar B^*B^*],[B_2^*\bar B^*B^*]\,\,\,\,\,\,\,\,\,$ & $[B_1\bar BB^*],[B_1\bar B^*B]\,\,\,\,\,\,\,\,\,$ & $[B_1\bar B^*B^*],[B_2^*\bar B^*B^*]$ \\
    \hline
    \end{tabular}
    \label{tab:loop}
\end{table}
Coupled-channel effects due to the coupling of bottomonium to open-bottom meson pairs can induce deviations from the above ratios~\cite{Guo:2014qra}.
On one hand, the $\Upsilon_D$ can couple to open-bottom meson pairs $B_1 \bar B^{(*)}$ and $B_2^*\bar B^*$ in $S$ waves.\footnote{Here, the meson pairs refer to the negative $C$-parity combinations, and we have neglected the charge conjugated pairs for simplicity of notation. They are included in calculations.}
On the other hand, the $S$-wave bottomonium couple to the same open-bottom meson pairs in $D$ waves~\cite{Eichten:1978tg, Li:2013yka}.
The difference can be understood by considering the conservation of the angular momentum of the light degrees of freedom, denoted by $s_\ell$, in the heavy quark limit.
Since there are no explicit light quarks inside bottomonia, $s_\ell$ is given purely by the orbital angular momentum, which is 0 and 2 for the $S$-wave and $D$-wave bottomonia, respectively.
The ground state $B$ and $B^*$ mesons (denoted as $H$) have $s_\ell^P=1/2^-$ with $P$ the parity of the bottom meson, while the narrow $B_1(5721)$ and $B_2^*(5747)$ resonances (denoted as $T$) have $s_\ell^P=3/2^+$. Conservation of $s_\ell$ thus requires the $\Upsilon_D$ to couple to the $B_1 \bar B^{(*)}$ and $B_2^*\bar B^*$ in $S$ waves, while the $\Upsilon_S$ couple to the same meson pairs in $D$ waves.
Since the mass of the $\Upsilon(10753)$ and $\Upsilon(10860)$ are only about 240~MeV and 130~MeV below the $B_1 \bar B$ threshold, the $S$-wave coupled-channel effects involving such open-bottom meson pairs can contribute significantly to the decays of the $\Upsilon_D$ component that we consider~\cite{Guo:2010ak}.
The coupled-channel effects are depicted by the triangle diagram shown in Fig.~\ref{fig:triangle_diagram}, and the heavy meson loops of interest are listed in Table~\ref{tab:loop}.

The significance of the triangle diagrams is evident from the NREFT power counting. 
Since the intermediate heavy mesons are treated nonrelativistically, the typical small parameter is the intermediate meson velocity $v$~\cite{Guo:2009wr,Guo:2010ak} (for a review, see Ref.~\cite{Guo:2017jvc}). 
As demonstrated in Ref.~\cite{Guo:2012tg}, for the triangle diagrams, this $v$ should be understood as the average of the two velocities in the corresponding unitary cuts depicted in Fig.~\ref{fig:triangle_diagram}. Its value is approximately 0.26 for $\Upsilon(10860)$ and 0.29 for $\Upsilon(10753)$ (for the explicit expression of estimating $v$, see Appendix~\ref{app:1}). 
For the initial state being a $D$-wave bottomonium, the power counting for the triangle diagram in Fig.~\ref{fig:triangle_diagram} is given by $\mathcal O(v^5/(v^2)^3)=\mathcal O(v^{-1})$, which indicates an enhancement for small $v$. Here, $v^5$ accounts for the nonrelativistic loop integral measure (each three-momentum component is of order $v$, and the temporal component is of order $v^2$), while $(v^2)^{-3}$ accounts for three nonrelativistic propagators.
If any of the bottomonium-bottom-meson vertices are of higher partial waves, there would be positive powers of $v$ from these vertices, and thus the corresponding triangle diagrams would be less important.
Therefore, we will only consider the triangle diagrams in Fig.~\ref{fig:triangle_diagram} for the decays of the $\Upsilon_D$ component, and consider the tree-level contribution in Fig.~\ref{fig:tree_diagram} for the decays of the $\Upsilon_S$ component.

The Lagrangian for vertices in such a triangle diagram is~\cite{Guo:2014qra,Fleming:2008yn,Guo:2013zbw} 

\begin{equation}
    \mathcal{L}_{\text{loop}}= \frac{g_4}{2}\left\langle\left(\bar T_a^{j\dagger}\sigma^iH_a^\dagger-\bar H_a^\dagger\sigma^iT_a^{j\dagger}\right)\Upsilon^{ij}\right\rangle+\frac{g_1}{2}\left\langle\chi^{i\dagger}H_a\sigma^i\bar H_a\right\rangle
    +\frac{c_\omega}{2}\left\langle H_a^\dagger T_a^i-\bar H_a^\dagger\bar T_a^i\right\rangle\omega^i+\text{h.c.},
    \label{loopEQ}
\end{equation}
where $H_a=\vec V_a\cdot\vec \sigma+P_a$ and $T_a^i=P^{ij}_{2,a}\sigma^j+\sqrt{\frac{2}{3}}P^i_{1,a}+i\sqrt{\frac{1}{6}}\epsilon^{ijk}P^j_{1,a}\sigma^k$
represent the $S$-wave bottom mesons with $s_\ell^P=1/2^-$ and the $P$-wave bottom mesons with $s_\ell^P=3/2^+$, respectively, with the subscript $a$ the light-flavor index. 
The fields $P_a$ and $V_a^i$ annihilate the pseudoscalar $B$ and vector $B^*$, respectively, while $P_{1,a}^i$ and $P_{2,a}^{ij}$ annihilate the axial-vector $B_1$ and tensor $B_2^*$ mesons, respectively. 
The fields for their anti-particles are $\bar H_a=-\vec{\bar{V_{a}}}\cdot\vec \sigma+\bar P_a$
and $\bar T_a^i=-\bar P^{ij}_{2,a}\sigma^j+\sqrt{\frac{2}{3}}\bar P^i_{1,a}-i\sqrt{\frac{1}{6}}\epsilon^{ijk}\bar P^j_{1,a}\sigma^k$ under the phase convention for the charge conjugation $A \overset{\mathcal C}{\longrightarrow} \bar A$, where $A$ represents any of the involved bottom mesons.

As the decays of $\Upsilon_D$ and $\Upsilon_S$ into $\chi_{bJ}\omega$ are saturated by triangle diagrams and tree diagrams respectively, the dependence of decay width ratios on $m_\Upsilon$ can be directly obtained as depicted in Fig.~\ref{PureSD}. 
The explicit expressions for loop amplitudes are detailed in Appendix~\ref{app:1}. Note that the involved scalar triangle loops are ultraviolet convergent, and thus no regularization is required. The deviation from Eq.~(\ref{1:3:5}) for the $S$-wave bottomonium near $\chi_{bJ}\omega$ thresholds is due to phase space effects (see the left panel of Fig.~\ref{PureSD}).
For the $D$-wave bottomonium, in addition to the phase space effect, the deviation is also due to the triangle diagram contributions since different intermediate states are involved (see the right panel of Fig.~\ref{PureSD}); if the $B_1$ and $B_2^*$ mesons have the same mass, and $B$ and $B^*$ mesons have the same mass, one would recover Eq.~\eqref{20:15:1}.
It is worth noting that the HQSS breaking effects from the triangle diagrams are amplified around the thresholds of $B_1\bar B$, $B_1\bar B^*$ and $B_2^*\bar B^*$. When the $B_1$ and $B_2^*$ widths are neglected, three distinct threshold cusps appear, which become smeared when the finite widths of $B_1$ and $B_2^*$ are taken into account.

\begin{figure}[tb]
	\centering
	\subfigure[]{
		\centering
		\includegraphics[width=7.9cm]{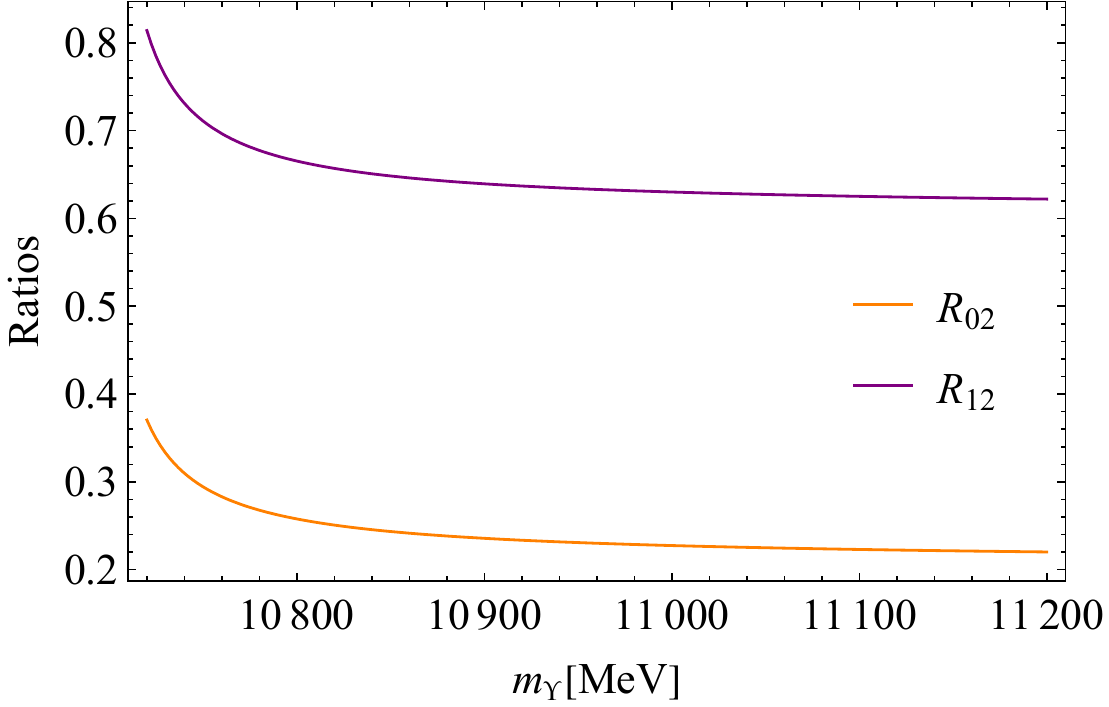}
	}
	\centering
	\subfigure[]{
		\centering
		\includegraphics[width=7.9cm]{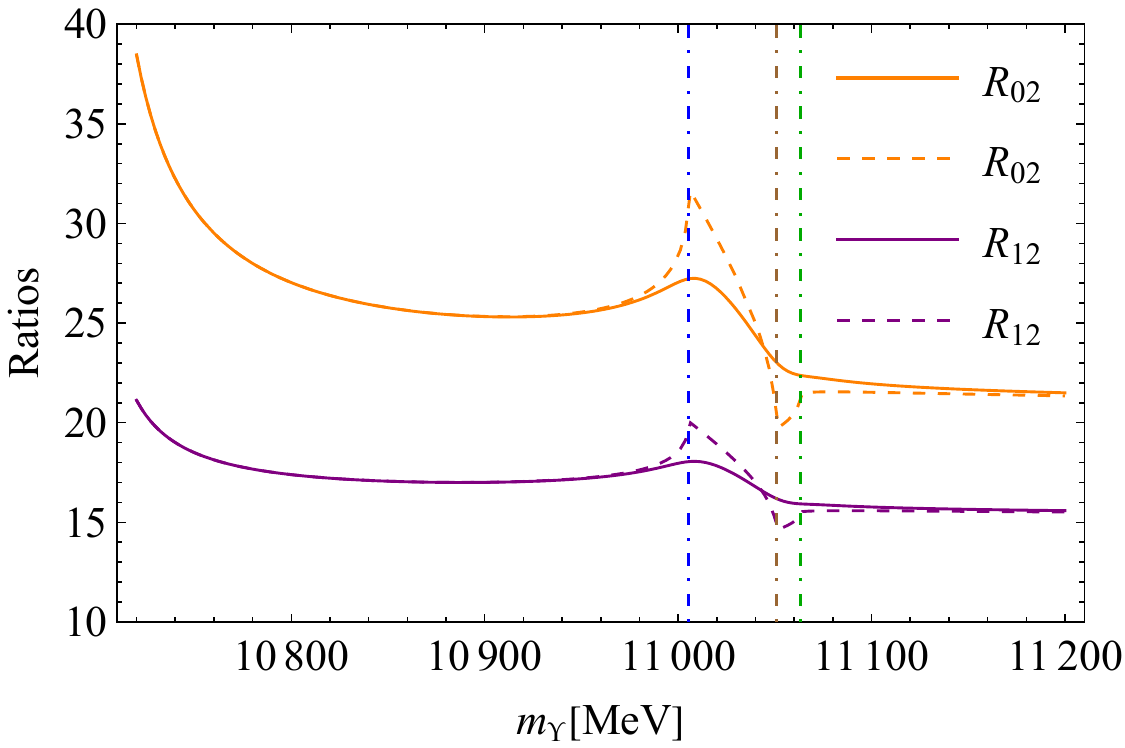}
	}
	\centering
	\caption{Dependence of the ratios $R_{02}$ and $R_{12}$ on the $\Upsilon$ mass: (a) for $S$-wave bottomonium decay, (b) for $D$-wave bottomonium decay. The vertical dot-dashed lines indicate the thresholds of $B_1 \bar B$, $B_1\bar B^*$ and $B_2^*\bar B^*$. In (b), the solid and dotted curves represent the results with and without considering the finite widths of $B_1$ and $B_2^*$, respectively. }
    \label{PureSD}
\end{figure}

At the mass of the $\Upsilon(10753)$, the branching fraction ratios in the right panel of Fig.~\ref{PureSD} read
\begin{equation}
\Gamma(\Upsilon_D\rightarrow\chi_{b0}\omega):\Gamma(\Upsilon_D\rightarrow\chi_{b1}\omega):\Gamma(\Upsilon_D\rightarrow\chi_{b2}\omega) \approx 29:18:1,
\end{equation}
which is drastically different from the experimental results in Eqs.~\eqref{10753r1} and \eqref{10753r2}.
Consequently, the $\Upsilon(10753)$ cannot be a pure $D$-wave bottomonium, and one way to reconcile the experimental data is to include an $S$-wave component.

In light of the aforementioned considerations, we consider $\Upsilon(10580)$ and $\Upsilon(10753)$ to be mixtures of $4S$ and $3D$ bottomonia ($\Upsilon_{4S}$ and $\Upsilon_{3D}$) with a mixing angle $\theta$:
\begin{align}
    \begin{pmatrix}
        \Upsilon(10580)\\
        \Upsilon(10753)
    \end{pmatrix}
    =\begin{pmatrix}
        \cos\theta &-\sin\theta\\
        \sin\theta &\,\cos\theta
    \end{pmatrix}
    \begin{pmatrix}
        \Upsilon_{4S}\\
        \Upsilon_{3D}
    \end{pmatrix},
\end{align}
and $\Upsilon(10860)$ to be the lower one of the $5S$-$4D$ ($\Upsilon_{5S}$-$\Upsilon_{4D}$) mixed states with a mixing angle $\theta'$ :
\begin{align}
    \begin{pmatrix}
        \Upsilon(10860)\\
        \Upsilon'(4D)
    \end{pmatrix}
    =\begin{pmatrix}
        \cos\theta' &-\sin\theta'\\
        \sin\theta' &\,\cos\theta'
    \end{pmatrix}
    \begin{pmatrix}
        \Upsilon_{5S}\\
        \Upsilon_{4D}
    \end{pmatrix},
\end{align}
while $\Upsilon'(4D)$ denotes the mixing partner of $\Upsilon(10860)$, which has not been experimentally observed yet. The mixing angles can be determined from the relation between the dielectron widths of these bottomonia and the wave functions at the origin~\cite{Badalian:2009bu}:
\begin{align}
    \Gamma_{ee}(\Upsilon(10580))&=\frac{4e_b^2\alpha^2}{m^2_{\Upsilon(10580)}}|R_{4S}(0)\cos\theta-R_{3D}(0)\sin\theta|^2\xi_1\beta_V,\label{eq:dielectron1}\\
    \Gamma_{ee}(\Upsilon(10753))&=\frac{4e_b^2\alpha^2}{m^2_{\Upsilon(10753)}}|R_{4S}(0)\sin\theta+R_{3D}(0)\cos\theta|^2\xi_1\beta_V,\label{eq:dielectron1'}\\
    \Gamma_{ee}(\Upsilon(10860))&=\frac{4e_b^2\alpha^2}{m^2_{\Upsilon(10860)}}|R_{5S}(0)\cos\theta'-R_{4D}(0)\sin\theta'|^2\xi_2\beta_V,\label{eq:dielectron2}\\
    \Gamma_{ee}(\Upsilon'(4D))&=\frac{4e_b^2\alpha^2}{m^2_{\Upsilon'(4D)}}|R_{5S}(0)\sin\theta'+R_{4D}(0)\cos\theta'|^2\xi_2\beta_V,\label{eq:dielectron2'}
\end{align}
where $e_b=-1/3$ is the charge of the $b$ quark, and $\alpha$ is the fine structure constant. $R_{nS}(0)$ and $R_{nD}(0)$ are the wave functions of $\Upsilon_{nS}$ and $\Upsilon_{nD}$ at the origin, listed in Table~\ref{tab:wf}.
\begin{table}[tb]
    \centering
    \caption{$\Upsilon$ wave functions at the origin in $4S$, $5S$, $3D$ and $4D$ states, extracted from Ref.~\cite{Badalian:2009bu}.}
    \begin{tabular}{lcccccccccc}
    \hline
    States & $4S$ & $5S$ & $3D$ & $4D$ \\
    \hline
    $R(0)$\,[$\text{GeV}^{3/2}$]\,\,\, & 1.506\,\,\, & 1.424\,\,\, & 0.0956\,\,\, & 0.107 \\
    \hline
    \end{tabular}
    \label{tab:wf}
\end{table}
$\xi_1=0.968$ and $\xi_2=0.966$ are relativistic factors and $\beta_V=0.80$ represents the QCD one-loop correction; these parameters are extracted from Ref.~\cite{Badalian:2009bu}. The mass shifts of the $5S$ and $4D$ states caused by mixing are described by\footnote{The formulas here differ from those in Ref.~\cite{Cheng:2011pb}. The difference arises because here we use the quark masses as input and derive the mixing in nonrelativistic quantum mechanics, whereas in Ref.~\cite{Cheng:2011pb} the mixing is derived in relativistic quantum field theory, in which the mass terms in the Lagrangian are quadratic. }
\begin{equation}
    m_{\Upsilon(10860)}=\frac{1}{2}\left(m_{\Upsilon(5S)}+m_{\Upsilon(4D)}-\sqrt{(m_{\Upsilon(5S)}-m_{\Upsilon(4D)})^2\sec^22\theta'}\right),
\end{equation}
\begin{equation}
    m_{\Upsilon'(4D)}=\frac{1}{2}\left(m_{\Upsilon(5S)}+m_{\Upsilon(4D)}+\sqrt{(m_{\Upsilon(5S)}-m_{\Upsilon(4D)})^2\sec^22\theta'}\right).
\end{equation}

The decay amplitudes for $\Upsilon(10753)$, $\Upsilon(10860)$, and $\Upsilon'(4D)$ into $\omega\chi_{bJ}$ are expressed as follows:
\begin{align}
    \mathcal A(\Upsilon(10753)\rightarrow\chi_{bJ}\omega)&=\mathcal A_{4SJ}\sin\theta+\mathcal A_{3DJ}\cos\theta, \\
    \mathcal A(\Upsilon(10860)\rightarrow\chi_{bJ}\omega)&=\mathcal A_{5SJ}\cos\theta'-\mathcal A_{4DJ}\sin\theta',\\
    \mathcal A(\Upsilon'(4D)\rightarrow\chi_{bJ}\omega)&=\mathcal A_{5SJ}\sin\theta'+\mathcal A_{4DJ}\cos\theta',
\end{align}
where $\mathcal A_{nSJ}$ denotes the decay amplitude from the $nS$-wave initial state to $\chi_{bJ}\omega$, containing only tree-level contributions, and $\mathcal A_{nDJ}$ denotes the decay amplitude from the $nD$-wave initial state to $\chi_{bJ}\omega$, containing only coupled-channel effect contributions from bottom-meson loops listed in Table~\ref{tab:loop}.

\section{Results and discussion}

\begin{figure}[tb]
    \centering
    \includegraphics[width=0.5\linewidth]{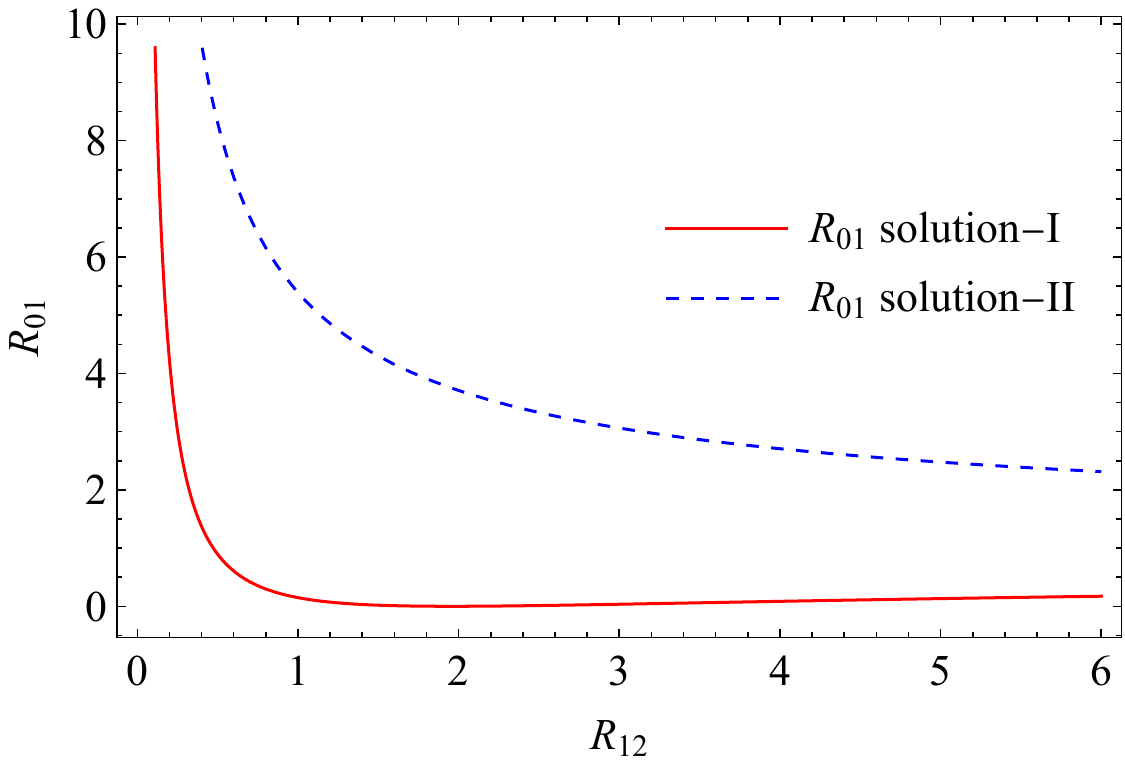}
    \caption{Dependence of ratio $R_{01}$ on ratio $R_{12}$ for the $4S$-$3D$ mixing of $\Upsilon(10753)$.}
    \label{fig:R01_R12}
\end{figure}
The ratios of decay amplitudes from $S$-wave and $D$-wave bottomonium components always appear together with a factor of $\tan\theta$, which can be eliminated to obtain a relation between the two ratios $R_{01}$ and $R_{12}$.
In the $4S$-$3D$ mixing scheme, such a relation is illustrated in Fig.~\ref{fig:R01_R12}, where two solutions for $R_{01}$ (denoted I and II) emerge from interference between $S$-wave and $D$-wave components.

\begin{table}[tb]
    \centering
    \caption{Branching fractions for $\Upsilon(10753) \to \chi_{bJ} \omega$ using inputs from Eq.~\eqref{10753r1}, in units of $10^{-3}$. For $\Upsilon(10753) \to \chi_{b0} \omega$, two values are given: the first corresponds to Solution-I for $R_{01}$, and the second to Solution-II.}
    \begin{tabular}{lcccc}
    \hline
    Decay modes & {\quad$\Upsilon(10753)\rightarrow\chi_{b0}\omega$\quad} &{$\Upsilon(10753)\rightarrow\chi_{b1}\omega$\quad} & {$\Upsilon(10753)\rightarrow\chi_{b2}\omega$\quad} \\
    \hline
    $\theta=(27\pm4)^\circ$ & $0.5^{+2.4}_{-0.5}$, $33\pm21$ & $6.8\pm5.0 $ & $5.7\pm5.3 $ \\
    \hline
    $\theta=(-35\pm4)^\circ$ & $0.5^{+2.3}_{-0.5}$,  $ 32\pm20$ & $6.3\pm4.8 $ & $5.3\pm5.1 $ \\
    \hline
    \end{tabular}
    \label{tab:10753Con}
\end{table}
\begin{table}[tb]
    \centering
    \caption{Branching fractions for $\Upsilon(10753) \to \chi_{bJ} \omega$ using inputs from Eq.~\eqref{10753r2}, in units of $10^{-3}$. For $\Upsilon(10753) \to \chi_{b0} \omega$, two values are given: the first corresponds to Solution-I for $R_{01}$, and the second to Solution-II.}
    \begin{tabular}{lcccc}
    \hline
    Decay modes & {\quad$\Upsilon(10753)\rightarrow\chi_{b0}\omega$\quad} &{$\Upsilon(10753)\rightarrow\chi_{b1}\omega$\quad} & {$\Upsilon(10753)\rightarrow\chi_{b2}\omega$\quad} \\
    \hline
    $\theta=(27\pm4)^\circ$ & $0.4^{+2.0}_{-0.4} $, $ 92\pm38$ & $22\pm10 $ & $14\pm8 $ \\
    \hline
    $\theta=(-35\pm4)^\circ$ & $0.3^{+1.9}_{-0.3} $,  $ 88\pm36$ & $21\pm10 $ & $14\pm8 $ \\
    \hline
    \end{tabular}
    \label{tab:10753De}
\end{table}

Using the measured dielectron width $\Gamma_{ee}(\Upsilon(10580)) = (0.272 \pm 0.029)$ keV~\cite{ParticleDataGroup:2024cfk} and Eq.~\eqref{eq:dielectron1}, we determine the mixing angle $\theta$ to be $(27 \pm 4)^\circ$ or $(-35 \pm 4)^\circ$, which is consistent with previous studies~\cite{Badalian:2009bu,Li:2021jjt}. 
These values yield estimates for the dielectron width of $\Upsilon(10753)$ as $\Gamma_{ee}(10753) = (93 \pm 22)$~eV for $\theta = (27 \pm 4)^\circ$ and $\Gamma_{ee}(10753) = (97 \pm 22)$~eV for $\theta = (-35 \pm 4)^\circ$. 
Consequently, the branching ratios for $\Upsilon(10753) \to \chi_{bJ} \omega$ are calculated from the Belle~II results for constructive interference (Eq.~\eqref{10753r1}) and destructive interference (Eq.~\eqref{10753r2}), as presented in Tables~\ref{tab:10753Con} and~\ref{tab:10753De}, respectively.
It is reasonable that the partial widths of $\Upsilon(10753) \to \chi_{bJ} \omega$, as Okubo-Zweig-Iizuka-suppressed processes, are at the order of $10$ to $100$~keV.\footnote{The total width of $\Upsilon(10753)$ is $36_{-12}^{+18}$~MeV~\cite{Belle:2019cbt}.} 
The experimental measurements, as given in Eqs.~\eqref{10753r1} and~\eqref{10753r2}, have relative uncertainties exceeding 50\%, which inevitably introduces significant uncertainties into our predictions.

\begin{figure}[t]
    \centering
    \includegraphics[width=9cm]{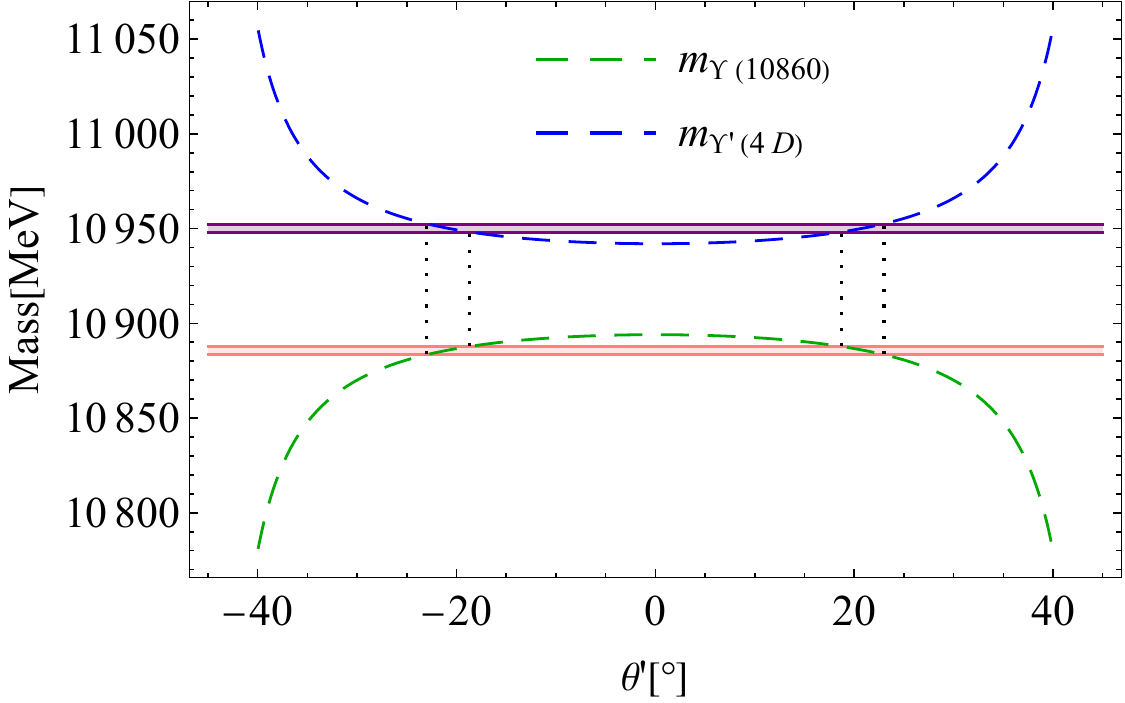}
    \caption{Dependence of the masses of $5S$-$4D$ mixed states on the mixing angle $\theta'$. The pink band represents the measured mass of $\Upsilon(10860)$, and the purple band shows the predicted mass of $\Upsilon'(4D)$. The pure $5S$ and $4D$ state masses are taken to be 10894~MeV and 10942~MeV, respectively, based on the quark model with a flattened potential in the non-relativistic limit~\cite{Badalian:2009bu}.}
    \label{fig:Mass_ThetaPrime}
\end{figure}
The $5S$-$4D$ mixing angle $\theta'$ can be constrained to lie in the range between $-33^\circ$ and $25^\circ$ using $\Gamma_{ee}(\Upsilon(10860)) = (0.31 \pm 0.07)$~keV and Eq.~\eqref{eq:dielectron2}. 
By analyzing the $\theta'$-dependence of the masses $m_{\Upsilon(10860)}$ and $m_{\Upsilon'(4D)}$, as shown in Fig.~\ref{fig:Mass_ThetaPrime}, $\theta'$ can be further constrained to $(21 \pm 2)^\circ$ or $(-21 \pm 2)^\circ$,

\footnote{Applying the same method to the $4S$-$3D$ mixing and using the masses from the flattened potential in the non-relativistic limit (10657 MeV and 10717 MeV for the $4S$ and $3D$ states, respectively) in Ref.~\cite{Badalian:2009bu} as input, the mixing angle $\theta$ is found to be about $\pm37^\circ$ from the $\Upsilon(10580)$ mass and $\pm31^\circ$ from the $\Upsilon(10753)$ mass, respectively. These values are in the same ballpark as those determined from the dielectron width of the $\Upsilon(10580)$.} 
which is consistent with the $\mathcal{O}(20^\circ)$ predictions in Ref.~\cite{Guo:2014qra}. 
The mass of $\Upsilon'(4D)$ is predicted to be approximately 10950 MeV (thus, $\Upsilon'(4D)$ will be denoted as $\Upsilon(10950)$ in the following), yielding dielectron widths of $\Gamma_{ee}(\Upsilon(10950)) = (57 \pm 8)$~eV for $\theta' = (21 \pm 2)^\circ$ and $(26 \pm 6)$~eV for $\theta' = (-21 \pm 2)^\circ$. These small dielectron widths explain why $\Upsilon(10950)$ has not been observed in experiments to date.

\begin{figure}[tb]
	\centering
	\subfigure[]{
		\centering
		\includegraphics[width=7.75cm]{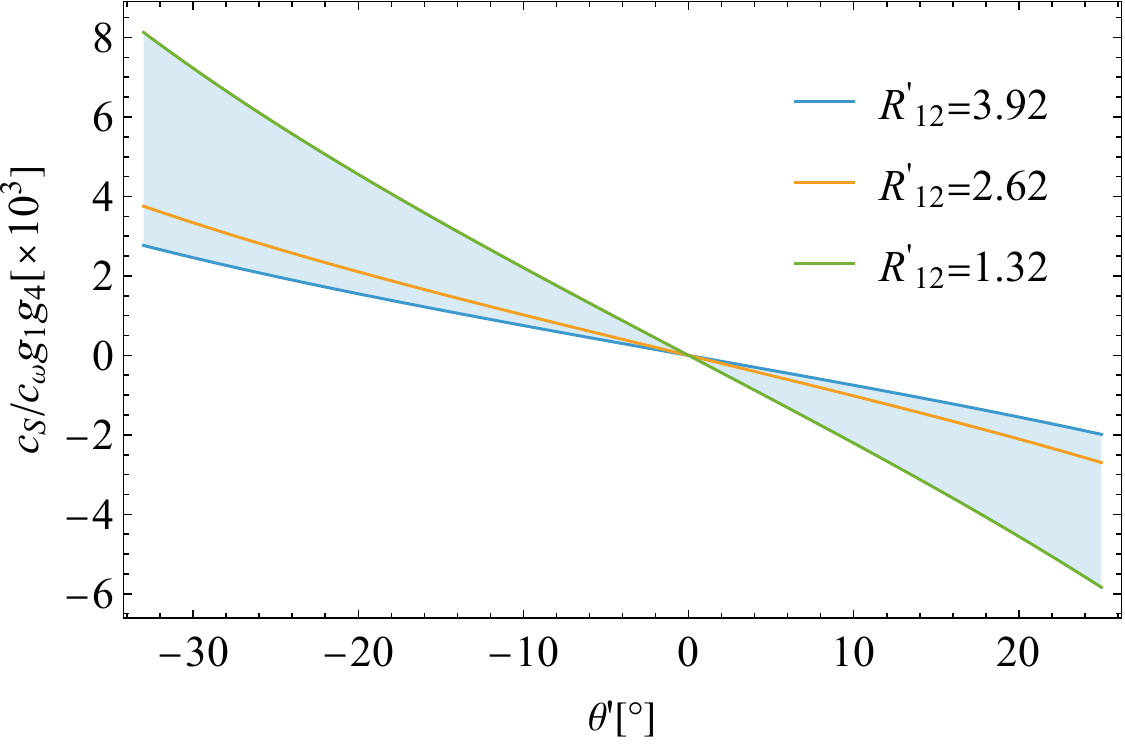}
	}
	\centering
	\subfigure[]{
		\centering
		\includegraphics[width=8.05cm]{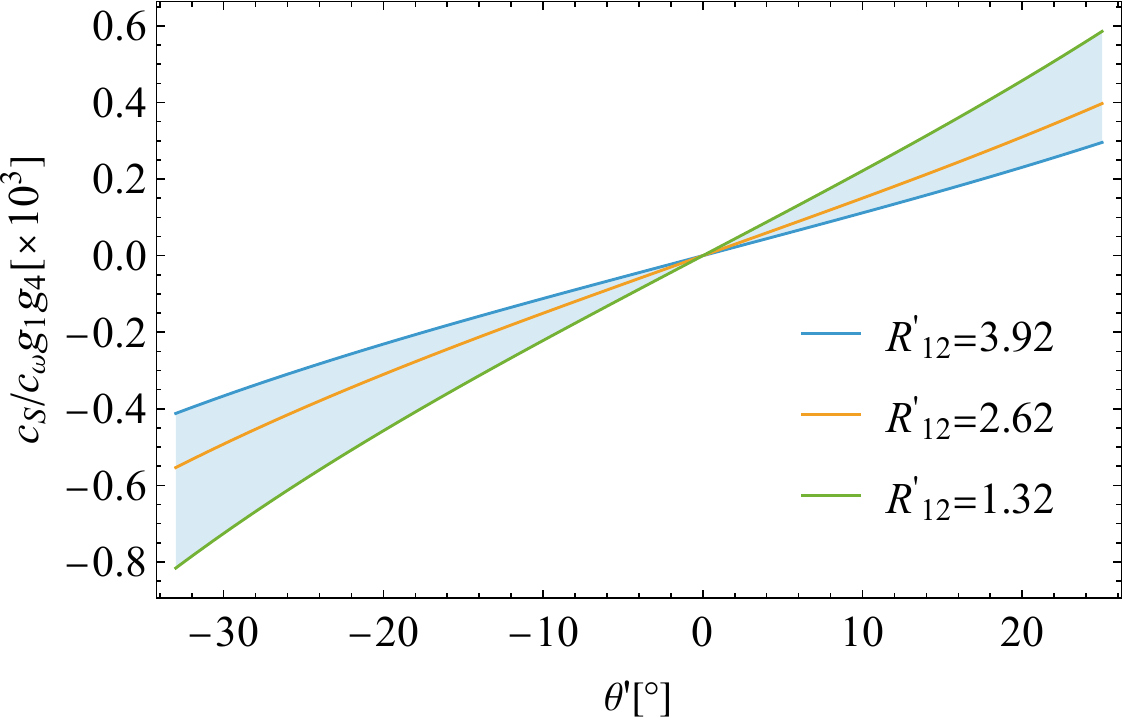}
	}
	\centering
	\caption{Dependence of the coupling constant ratio $c_S / (c_\omega g_1 g_4)$ on the $5S$-$4D$ mixing angle $\theta'$, with $R'_{12}$ fixed at its upper limit, central value, and lower limit. (a) and (b) show two possible solutions arising from the $S$-$D$ interference.}
    \label{fig:cscg}
\end{figure}

There are two solutions for the coupling constant ratio $c_S / (c_\omega g_1 g_4)$ as a function of $\theta'$ for a given $R'_{12}$, which also arise from the interference between $S$-wave and $D$-wave components, as illustrated in Fig.~\ref{fig:cscg}. 
By substituting the derived $c_S / (c_\omega g_1 g_4)$ into the partial decay widths of $\Upsilon(10860)$, as given by Eq.~\eqref{10860r}, we can determine the absolute values of the coupling constants $c_\omega g_1 g_4$ and $c_S$. Using these values and the total width $\Gamma(\Upsilon(10860)) = (37 \pm 4)$~MeV~\cite{ParticleDataGroup:2024cfk,Belle:2015tbu,Belle:2019cbt}, the partial decay widths for $\Upsilon(10950) \to \chi_{bJ} \omega$ are predicted, as presented in Table~\ref{tab:4D}.
\begin{table}[tb]
    \centering
    \caption{Predictions for the decay widths of $\Upsilon(10950)\rightarrow\chi_{bJ}\omega$ with $\theta'=21^\circ\pm2^\circ$ and $\theta'=-21^\circ\pm2^\circ$, in units of MeV. Both values of $\theta'$ yield identical predictions for the $\Upsilon(10950)\rightarrow\chi_{bJ}\omega$ decay widths. Solution-I corresponds to the left panel of Fig.~\ref{fig:cscg}, while Solution-II corresponds to the right panel.}
    \begin{tabular}{lcccc}
    \hline
    Decay modes & {\quad$\Upsilon(10950)\rightarrow\chi_{b0}\omega$\quad} &{$\Upsilon(10950)\rightarrow\chi_{b1}\omega$\quad} & {$\Upsilon(10950)\rightarrow\chi_{b2}\omega$\quad} \\
    \hline
    Solution-I & $0.18\pm0.10$ & $0.068\pm0.056 $ & $0.023\pm0.006 $ \\
    \hline
    Solution-II & $1.3\pm0.4 $ & $0.94\pm0.29 $ & $0.038\pm0.013 $ \\
    \hline
    \end{tabular}
    \label{tab:4D}
\end{table}

\section{Summary}

In summary, we model $\Upsilon(10580)$ and $\Upsilon(10753)$ as $4S$-$3D$ bottomonium mixtures and $\Upsilon(10860)$ as a $5S$-$4D$ mixed state, predicting the existence of its partner $\Upsilon(10950)$ with a large $D$-wave component. The mixing angles are determined from dielectron decay widths and mass shifts: $\theta = (27 \pm 4)^\circ$ or $(-35 \pm 4)^\circ$ for $4S$-$3D$ mixing, and $\theta' = \pm (21 \pm 2)^\circ$ for $5S$-$4D$ mixing. 
Using a non-relativistic effective field theory that incorporates HQSS, we investigate the decays of $\Upsilon(10753)$ and $\Upsilon(10860)$ into $\omega\chi_{bJ}$. The $S$-wave component decay is dominated by tree-level processes at hadronic level, while the $D$-wave component is primarily influenced by coupled-channel effects. The Belle \cite{Belle:2014sys} and Belle~II \cite{Belle-II:2022xdi} results highlight the necessity of the $S$-$D$ mixing scheme for both states. We subsequently predict the dielectron widths and the partial decay widths into $\chi_{bJ} \omega$ for $\Upsilon(10753)$ and $\Upsilon(10950)$. 
$\Upsilon(10950)$ is predicted to have a small dielectron width of $(57 \pm 8)$~eV for $\theta' = (21 \pm 2)^\circ$ or $(26 \pm 6)$~eV for $\theta' = (-21 \pm 2)^\circ$. Such a small dielectron width hinders its observation directly in $e^+e^-$ collisions. 
Note that the mixing angles were determined using quark model predictions for the masses and wave functions at the origin, whose systematic uncertainties are difficult to quantify. Therefore, these values should be regarded as reasonable estimates rather than precise predictions.
These results provide valuable insights into bottomonium spectroscopy and can be tested through more precise measurements of $\Upsilon(10753)$ and $\Upsilon(10860)$ decay properties.

\begin{acknowledgments}
We are grateful to Cheng-Ping Shen for useful discussions. This work is supported in part by the National Natural Science Foundation of China (NSFC) under Grants No. 12361141819, No. 12125507,  and No. 12447101; by the National Key R\&D Program of China under Grant No. 2023YFA1606703; and by the Chinese Academy of Sciences under Grant No. YSBR-101. 
\end{acknowledgments}

\appendix
\section{Decay Amplitudes}\label{app:1}
The decay amplitudes are listed here. 

In the tree level, the amplitudes are
\begin{equation}
    i\mathcal A_{S0}^{\text{tree}}=\frac{i}{\sqrt{3}}N_0c_S\epsilon_\Upsilon^i\epsilon_\omega^i,\quad i\mathcal A_{S1}^{\text{tree}}=\frac{i}{\sqrt{2}}N_1c_S\epsilon_{ijk}\epsilon_{\chi_1}^i\epsilon_\Upsilon^j\epsilon_\omega^k,\quad i\mathcal A_{S2}^{\text{tree}}=-iN_2c_S\epsilon_{\chi_2}^{ij}\epsilon_\Upsilon^i\epsilon_\omega^j,
\end{equation}
and
\begin{equation}
    i\mathcal A_{D0}^{\text{tree}}=\frac{i\sqrt{5}}{3}N_0c_D\epsilon_\Upsilon^i\epsilon_\omega^i,\quad i\mathcal A_{D1}^{\text{tree}}=-\frac{i}{2}\sqrt{\frac{5}{6}}N_1c_D\epsilon_{ijk}\epsilon_{\chi_1}^i\epsilon_\Upsilon^j\epsilon_\omega^k,\quad i\mathcal A_{D2}^{\text{tree}}=-\frac{i}{2\sqrt{15}}N_2c_D\epsilon_{\chi_2}^{ij}\epsilon_\Upsilon^i\epsilon_\omega^j.
\end{equation}
$\mathcal A_{SJ}^{\text{tree}}$ represents the amplitude of $\Upsilon(5S)\rightarrow \chi_{bJ}\omega$ at tree level, where $\epsilon_\Upsilon^i$ is the polarization vector of $\Upsilon$ and $\epsilon_{\chi_2}^{ij}$ is the polarization tensor of $\chi_{b2}$. The factors $N_J=\sqrt{2m_{\Upsilon}\cdot 2m_{\chi_{bJ}}}\quad(J=0,1,2)$ account for the nonrelativistic normalization, involving the masses of heavy particles in the initial and final states. 

The triangle loops only contribute to the $D$-wave initial state, and the corresponding amplitudes are 
\begin{align}
    i\mathcal A_{D0}^{\text{loop}}&=-\frac{i\sqrt{5}}{3}N_0c_\omega g_1g_4\epsilon_\Upsilon^i\epsilon_\omega^i[6I(m_{B_1},m_B,m_B,q_\omega)+I(m_{B_1},m_{B^*},m_{B^*},q_\omega)+I(m_{B_2^*},m_{B^*},m_{B^*},q_\omega)], \notag \\
    i\mathcal A_{D1}^{\text{loop}}&=i\sqrt{\frac{10}{3}}N_1c_\omega g_1g_4\epsilon_{ijk}\epsilon_{\chi_1}^i\epsilon_\Upsilon^j\epsilon_\omega^k[I(m_{B_1},m_{B},m_{B^*},q_\omega)+I(m_{B_1},m_{B^*},m_{B},q_\omega)],\\
    i\mathcal A_{D2}^{\text{tree}}&=\frac{i}{\sqrt{15}}N_2c_\omega g_1g_4\epsilon_{\chi_2}^{ij}\epsilon_\Upsilon^i\epsilon_\omega^j[5I(m_{B_1},m_{B^*},m_{B^*},q_\omega)-I(m_{B_2^*},m_{B^*},m_{B^*},q_\omega)]. \notag
\end{align}
In the loop amplitudes, 
\begin{dmath}
    I(m_1,m_2,m_3,q_\omega)=i\int\frac{d^4l}{(2\pi)^4}\frac{1}{(l^0-\omega_1(\vec l)+i\Gamma_1/2)(m_\Upsilon-l^0-\omega_2(-\vec l)+i\epsilon)(l^0-q_\omega^0-\omega_3(\vec l-\vec q_\omega)+i\epsilon)}
\end{dmath}
is a nonrelativistic scalar triangle loop integral evaluated at the rest frame of the initial state, with $\omega_i(\vec k)=m_i+\vec k^2/(2m_i)$, and it can be analytically expressed as \cite{Guo:2010ak,Guo:2013zbw,Guo:2014qra}: 
\begin{equation}
    I(m_1,m_2,m_3,q)=\frac{\mu_{12}\mu_{23}}{2\pi\sqrt{a}}\left(\arctan\frac{c'-c}{2\sqrt{ac}}+\arctan\frac{2a+c-c'}{2\sqrt{a(c'-a)}}\right),
\end{equation}
with
$$a=(\frac{\mu_{23}}{m_3}\vec q_\omega)^2,\quad c=2\mu_{12}b_{12},\quad c'=2\mu_{23}b_{23}+\frac{\mu_{23}}{m_3}(\vec q_\omega)^2,$$
$$\mu_{ij}=\frac{m_im_j}{m_i+m_j},\quad b_{12}=m_1-\frac{i\Gamma_1}{2}+m_2-m_\Upsilon,\quad b_{23}=m_2+m_3+q_\omega^0-m_\Upsilon,$$
where $q_\omega$ is the 4-momentum of $\omega$, and $m_\Upsilon$ should take the value of the mass of the corresponding initial state.

$\sqrt{|c|}$ and $\sqrt{|c'|}$ provide two momentum scales for intermediate mesons. The velocities corresponding to the respective unitary cuts can be defined as $v_1=\sqrt{|c|}/(2\mu_{12})$ and $v_2 = \sqrt{|c'-a|}/(2\mu_{23})$. Therefore, the average velocity is given by $v = (v_1+v_2)/2 \approx 0.26$ for $\Upsilon(10860)$~\cite{Guo:2014qra} and $0.29$ for $\Upsilon(10753)$.

\bibliography{refs}

\begin{thebibliography}{38}%
\makeatletter
\providecommand \@ifxundefined [1]{%
 \@ifx{#1\undefined}
}%
\providecommand \@ifnum [1]{%
 \ifnum #1\expandafter \@firstoftwo
 \else \expandafter \@secondoftwo
 \fi
}%
\providecommand \@ifx [1]{%
 \ifx #1\expandafter \@firstoftwo
 \else \expandafter \@secondoftwo
 \fi
}%
\providecommand \natexlab [1]{#1}%
\providecommand \enquote  [1]{``#1''}%
\providecommand \bibnamefont  [1]{#1}%
\providecommand \bibfnamefont [1]{#1}%
\providecommand \citenamefont [1]{#1}%
\providecommand \href@noop [0]{\@secondoftwo}%
\providecommand \href [0]{\begingroup \@sanitize@url \@href}%
\providecommand \@href[1]{\@@startlink{#1}\@@href}%
\providecommand \@@href[1]{\endgroup#1\@@endlink}%
\providecommand \@sanitize@url [0]{\catcode `\\12\catcode `\$12\catcode `\&12\catcode `\#12\catcode `\^12\catcode `\_12\catcode `\%12\relax}%
\providecommand \@@startlink[1]{}%
\providecommand \@@endlink[0]{}%
\providecommand \url  [0]{\begingroup\@sanitize@url \@url }%
\providecommand \@url [1]{\endgroup\@href {#1}{\urlprefix }}%
\providecommand \urlprefix  [0]{URL }%
\providecommand \Eprint [0]{\href }%
\providecommand \doibase [0]{https://doi.org/}%
\providecommand \selectlanguage [0]{\@gobble}%
\providecommand \bibinfo  [0]{\@secondoftwo}%
\providecommand \bibfield  [0]{\@secondoftwo}%
\providecommand \translation [1]{[#1]}%
\providecommand \BibitemOpen [0]{}%
\providecommand \bibitemStop [0]{}%
\providecommand \bibitemNoStop [0]{.\EOS\space}%
\providecommand \EOS [0]{\spacefactor3000\relax}%
\providecommand \BibitemShut  [1]{\csname bibitem#1\endcsname}%
\let\auto@bib@innerbib\@empty
\bibitem [{\citenamefont {Herb}\ \emph {et~al.}(1977)\citenamefont {Herb} \emph {et~al.}}]{E288:1977xhf}%
  \BibitemOpen
  \bibfield  {author} {\bibinfo {author} {\bibfnamefont {S.~W.}\ \bibnamefont {Herb}} \emph {et~al.} (\bibinfo {collaboration} {E288}),\ }\bibfield  {title} {\bibinfo {title} {{Observation of a Dimuon Resonance at 9.5 GeV in 400-GeV Proton-Nucleus Collisions}},\ }\href {https://doi.org/10.1103/PhysRevLett.39.252} {\bibfield  {journal} {\bibinfo  {journal} {Phys. Rev. Lett.}\ }\textbf {\bibinfo {volume} {39}},\ \bibinfo {pages} {252} (\bibinfo {year} {1977})}\BibitemShut {NoStop}%
\bibitem [{\citenamefont {Innes}\ \emph {et~al.}(1977)\citenamefont {Innes} \emph {et~al.}}]{E288:1977efs}%
  \BibitemOpen
  \bibfield  {author} {\bibinfo {author} {\bibfnamefont {W.~R.}\ \bibnamefont {Innes}} \emph {et~al.} (\bibinfo {collaboration} {E288}),\ }\bibfield  {title} {\bibinfo {title} {{Observation of Structure in the $\Upsilon$ Region}},\ }\href {https://doi.org/10.1103/PhysRevLett.39.1240} {\bibfield  {journal} {\bibinfo  {journal} {Phys. Rev. Lett.}\ }\textbf {\bibinfo {volume} {39}},\ \bibinfo {pages} {1240} (\bibinfo {year} {1977})},\ \bibinfo {note} {[Erratum: Phys.Rev.Lett. 39, 1640 (1977)]}\BibitemShut {NoStop}%
\bibitem [{\citenamefont {Bonvicini}\ \emph {et~al.}(2004)\citenamefont {Bonvicini} \emph {et~al.}}]{CLEO:2004npj}%
  \BibitemOpen
  \bibfield  {author} {\bibinfo {author} {\bibfnamefont {G.}~\bibnamefont {Bonvicini}} \emph {et~al.} (\bibinfo {collaboration} {CLEO}),\ }\bibfield  {title} {\bibinfo {title} {First observation of a {$\Upsilon(1D)$} state},\ }\href {https://doi.org/10.1103/PhysRevD.70.032001} {\bibfield  {journal} {\bibinfo  {journal} {Phys. Rev. D}\ }\textbf {\bibinfo {volume} {70}},\ \bibinfo {pages} {032001} (\bibinfo {year} {2004})},\ \Eprint {https://arxiv.org/abs/hep-ex/0404021} {arXiv:hep-ex/0404021} \BibitemShut {NoStop}%
\bibitem [{\citenamefont {{del Amo Sanchez}}\ \emph {et~al.}(2010)\citenamefont {{del Amo Sanchez}} \emph {et~al.}}]{BaBar:2010tqb}%
  \BibitemOpen
  \bibfield  {author} {\bibinfo {author} {\bibfnamefont {P.}~\bibnamefont {{del Amo Sanchez}}} \emph {et~al.} (\bibinfo {collaboration} {BaBar}),\ }\bibfield  {title} {\bibinfo {title} {Observation of the ${{\Upsilon}}({{1^3D}}_{{J}})$ bottomonium state through decays to $\pi^+\pi^-{{\Upsilon (1S)}}$},\ }\href {https://doi.org/10.1103/PhysRevD.82.111102} {\bibfield  {journal} {\bibinfo  {journal} {Phys. Rev. D}\ }\textbf {\bibinfo {volume} {82}},\ \bibinfo {pages} {111102} (\bibinfo {year} {2010})},\ \Eprint {https://arxiv.org/abs/1004.0175} {arXiv:1004.0175 [hep-ex]} \BibitemShut {NoStop}%
\bibitem [{\citenamefont {Godfrey}\ and\ \citenamefont {Isgur}(1985)}]{Godfrey:1985xj}%
  \BibitemOpen
  \bibfield  {author} {\bibinfo {author} {\bibfnamefont {S.}~\bibnamefont {Godfrey}}\ and\ \bibinfo {author} {\bibfnamefont {N.}~\bibnamefont {Isgur}},\ }\bibfield  {title} {\bibinfo {title} {Mesons in a relativized quark model with chromodynamics},\ }\href {https://doi.org/10.1103/PhysRevD.32.189} {\bibfield  {journal} {\bibinfo  {journal} {Physical Review D}\ }\textbf {\bibinfo {volume} {32}},\ \bibinfo {pages} {189} (\bibinfo {year} {1985})}\BibitemShut {NoStop}%
\bibitem [{\citenamefont {Brambilla}\ \emph {et~al.}(2011)\citenamefont {Brambilla} \emph {et~al.}}]{Brambilla:2010cs}%
  \BibitemOpen
  \bibfield  {author} {\bibinfo {author} {\bibfnamefont {N.}~\bibnamefont {Brambilla}} \emph {et~al.},\ }\bibfield  {title} {\bibinfo {title} {Heavy quarkonium: Progress, puzzles, and opportunities},\ }\href {https://doi.org/10.1140/epjc/s10052-010-1534-9} {\bibfield  {journal} {\bibinfo  {journal} {Eur. Phys. J. C}\ }\textbf {\bibinfo {volume} {71}},\ \bibinfo {pages} {1534} (\bibinfo {year} {2011})},\ \Eprint {https://arxiv.org/abs/1010.5827} {arXiv:1010.5827 [hep-ph]} \BibitemShut {NoStop}%
\bibitem [{\citenamefont {Drutskoy}\ \emph {et~al.}(2013)\citenamefont {Drutskoy}, \citenamefont {Guo}, \citenamefont {{Llanes-Estrada}}, \citenamefont {Nefediev},\ and\ \citenamefont {{Torres-Rincon}}}]{Drutskoy:2012gt}%
  \BibitemOpen
  \bibfield  {author} {\bibinfo {author} {\bibfnamefont {A.~G.}\ \bibnamefont {Drutskoy}}, \bibinfo {author} {\bibfnamefont {F.-K.}\ \bibnamefont {Guo}}, \bibinfo {author} {\bibfnamefont {F.~J.}\ \bibnamefont {{Llanes-Estrada}}}, \bibinfo {author} {\bibfnamefont {A.~V.}\ \bibnamefont {Nefediev}},\ and\ \bibinfo {author} {\bibfnamefont {J.~M.}\ \bibnamefont {{Torres-Rincon}}},\ }\bibfield  {title} {\bibinfo {title} {Hadron physics potential of future high-luminosity {$B$-factories} at the {$\Upsilon(5S)$} and above},\ }\href {https://doi.org/10.1140/epja/i2013-13007-x} {\bibfield  {journal} {\bibinfo  {journal} {Eur. Phys. J. A}\ }\textbf {\bibinfo {volume} {49}},\ \bibinfo {pages} {7} (\bibinfo {year} {2013})},\ \Eprint {https://arxiv.org/abs/1210.6623} {arXiv:1210.6623 [hep-ph]} \BibitemShut {NoStop}%
\bibitem [{\citenamefont {Mizuk}\ \emph {et~al.}(2019)\citenamefont {Mizuk} \emph {et~al.}}]{Belle:2019cbt}%
  \BibitemOpen
  \bibfield  {author} {\bibinfo {author} {\bibfnamefont {R.}~\bibnamefont {Mizuk}} \emph {et~al.} (\bibinfo {collaboration} {Belle}),\ }\bibfield  {title} {\bibinfo {title} {{Observation of a new structure near 10.75 GeV in the energy dependence of the $e^{+} e^{-} \rightarrow \Upsilon(n S) \pi^{+} \pi^{-}(n=1,2,3)$ cross sections}},\ }\href {https://doi.org/10.1007/JHEP10(2019)220} {\bibfield  {journal} {\bibinfo  {journal} {JHEP}\ }\textbf {\bibinfo {volume} {10}},\ \bibinfo {pages} {220}},\ \Eprint {https://arxiv.org/abs/1905.05521} {arXiv:1905.05521 [hep-ex]} \BibitemShut {NoStop}%
\bibitem [{\citenamefont {Navas}\ and\ \citenamefont {{Others}}(2024)}]{ParticleDataGroup:2024cfk}%
  \BibitemOpen
  \bibfield  {author} {\bibinfo {author} {\bibfnamefont {S.}~\bibnamefont {Navas}}\ and\ \bibinfo {author} {\bibnamefont {{Others}}} (\bibinfo {collaboration} {Particle Data Group}),\ }\bibfield  {title} {\bibinfo {title} {Review of {{Particle Physics}}},\ }\href {https://doi.org/10.1103/PhysRevD.110.030001} {\bibfield  {journal} {\bibinfo  {journal} {Phys. Rev. D}\ }\textbf {\bibinfo {volume} {110}},\ \bibinfo {pages} {030001} (\bibinfo {year} {2024})}\BibitemShut {NoStop}%
\bibitem [{\citenamefont {Segovia}\ \emph {et~al.}(2016)\citenamefont {Segovia}, \citenamefont {Ortega}, \citenamefont {Entem},\ and\ \citenamefont {Fern{\'a}ndez}}]{Segovia:2016xqb}%
  \BibitemOpen
  \bibfield  {author} {\bibinfo {author} {\bibfnamefont {J.}~\bibnamefont {Segovia}}, \bibinfo {author} {\bibfnamefont {P.~G.}\ \bibnamefont {Ortega}}, \bibinfo {author} {\bibfnamefont {D.~R.}\ \bibnamefont {Entem}},\ and\ \bibinfo {author} {\bibfnamefont {F.}~\bibnamefont {Fern{\'a}ndez}},\ }\bibfield  {title} {\bibinfo {title} {Bottomonium spectrum revisited},\ }\href {https://doi.org/10.1103/PhysRevD.93.074027} {\bibfield  {journal} {\bibinfo  {journal} {Phys. Rev. D}\ }\textbf {\bibinfo {volume} {93}},\ \bibinfo {pages} {074027} (\bibinfo {year} {2016})},\ \Eprint {https://arxiv.org/abs/1601.05093} {arXiv:1601.05093 [hep-ph]} \BibitemShut {NoStop}%
\bibitem [{\citenamefont {Wang}\ \emph {et~al.}(2018)\citenamefont {Wang}, \citenamefont {Sun}, \citenamefont {Liu},\ and\ \citenamefont {Matsuki}}]{Wang:2018rjg}%
  \BibitemOpen
  \bibfield  {author} {\bibinfo {author} {\bibfnamefont {J.-Z.}\ \bibnamefont {Wang}}, \bibinfo {author} {\bibfnamefont {Z.-F.}\ \bibnamefont {Sun}}, \bibinfo {author} {\bibfnamefont {X.}~\bibnamefont {Liu}},\ and\ \bibinfo {author} {\bibfnamefont {T.}~\bibnamefont {Matsuki}},\ }\bibfield  {title} {\bibinfo {title} {{Higher bottomonium zoo}},\ }\href {https://doi.org/10.1140/epjc/s10052-018-6372-1} {\bibfield  {journal} {\bibinfo  {journal} {Eur. Phys. J. C}\ }\textbf {\bibinfo {volume} {78}},\ \bibinfo {pages} {915} (\bibinfo {year} {2018})},\ \Eprint {https://arxiv.org/abs/1802.04938} {arXiv:1802.04938 [hep-ph]} \BibitemShut {NoStop}%
\bibitem [{\citenamefont {Godfrey}\ and\ \citenamefont {Moats}(2015)}]{Godfrey:2015dia}%
  \BibitemOpen
  \bibfield  {author} {\bibinfo {author} {\bibfnamefont {S.}~\bibnamefont {Godfrey}}\ and\ \bibinfo {author} {\bibfnamefont {K.}~\bibnamefont {Moats}},\ }\bibfield  {title} {\bibinfo {title} {{Bottomonium Mesons and Strategies for their Observation}},\ }\href {https://doi.org/10.1103/PhysRevD.92.054034} {\bibfield  {journal} {\bibinfo  {journal} {Phys. Rev. D}\ }\textbf {\bibinfo {volume} {92}},\ \bibinfo {pages} {054034} (\bibinfo {year} {2015})},\ \Eprint {https://arxiv.org/abs/1507.00024} {arXiv:1507.00024 [hep-ph]} \BibitemShut {NoStop}%
\bibitem [{\citenamefont {Li}\ \emph {et~al.}(2021)\citenamefont {Li}, \citenamefont {Bai}, \citenamefont {Huang},\ and\ \citenamefont {Liu}}]{Li:2021jjt}%
  \BibitemOpen
  \bibfield  {author} {\bibinfo {author} {\bibfnamefont {Y.-S.}\ \bibnamefont {Li}}, \bibinfo {author} {\bibfnamefont {Z.-Y.}\ \bibnamefont {Bai}}, \bibinfo {author} {\bibfnamefont {Q.}~\bibnamefont {Huang}},\ and\ \bibinfo {author} {\bibfnamefont {X.}~\bibnamefont {Liu}},\ }\bibfield  {title} {\bibinfo {title} {Hidden-bottom hadronic decays of ${{\Upsilon}}(10753)$ with a $\eta^{(')}$ or $\omega$ emission},\ }\href {https://doi.org/10.1103/PhysRevD.104.034036} {\bibfield  {journal} {\bibinfo  {journal} {Phys. Rev. D}\ }\textbf {\bibinfo {volume} {104}},\ \bibinfo {pages} {034036} (\bibinfo {year} {2021})},\ \Eprint {https://arxiv.org/abs/2106.14123} {arXiv:2106.14123 [hep-ph]} \BibitemShut {NoStop}%
\bibitem [{\citenamefont {Bai}\ \emph {et~al.}(2022)\citenamefont {Bai}, \citenamefont {Li}, \citenamefont {Huang}, \citenamefont {Liu},\ and\ \citenamefont {Matsuki}}]{Bai:2022cfz}%
  \BibitemOpen
  \bibfield  {author} {\bibinfo {author} {\bibfnamefont {Z.-Y.}\ \bibnamefont {Bai}}, \bibinfo {author} {\bibfnamefont {Y.-S.}\ \bibnamefont {Li}}, \bibinfo {author} {\bibfnamefont {Q.}~\bibnamefont {Huang}}, \bibinfo {author} {\bibfnamefont {X.}~\bibnamefont {Liu}},\ and\ \bibinfo {author} {\bibfnamefont {T.}~\bibnamefont {Matsuki}},\ }\bibfield  {title} {\bibinfo {title} {${{\Upsilon(10753)}}\rightarrow{{\Upsilon}}({{nS}})\pi^+\pi^-$ decays induced by hadronic loop mechanism},\ }\href {https://doi.org/10.1103/PhysRevD.105.074007} {\bibfield  {journal} {\bibinfo  {journal} {Phys. Rev. D}\ }\textbf {\bibinfo {volume} {105}},\ \bibinfo {pages} {074007} (\bibinfo {year} {2022})},\ \Eprint {https://arxiv.org/abs/2201.12715} {arXiv:2201.12715 [hep-ph]} \BibitemShut {NoStop}%
\bibitem [{\citenamefont {Tarr{\'u}s~Castell{\`a}}\ and\ \citenamefont {Passemar}(2021)}]{TarrusCastella:2021pld}%
  \BibitemOpen
  \bibfield  {author} {\bibinfo {author} {\bibfnamefont {J.}~\bibnamefont {Tarr{\'u}s~Castell{\`a}}}\ and\ \bibinfo {author} {\bibfnamefont {E.}~\bibnamefont {Passemar}},\ }\bibfield  {title} {\bibinfo {title} {Exotic to standard bottomonium transitions},\ }\href {https://doi.org/10.1103/PhysRevD.104.034019} {\bibfield  {journal} {\bibinfo  {journal} {Phys. Rev. D}\ }\textbf {\bibinfo {volume} {104}},\ \bibinfo {pages} {034019} (\bibinfo {year} {2021})},\ \Eprint {https://arxiv.org/abs/2104.03975} {arXiv:2104.03975 [hep-ph]} \BibitemShut {NoStop}%
\bibitem [{\citenamefont {Wang}(2019)}]{Wang:2019veq}%
  \BibitemOpen
  \bibfield  {author} {\bibinfo {author} {\bibfnamefont {Z.-G.}\ \bibnamefont {Wang}},\ }\bibfield  {title} {\bibinfo {title} {{Vector hidden-bottom tetraquark candidate: $Y(10750)$}},\ }\href {https://doi.org/10.1088/1674-1137/43/12/123102} {\bibfield  {journal} {\bibinfo  {journal} {Chin. Phys. C}\ }\textbf {\bibinfo {volume} {43}},\ \bibinfo {pages} {123102} (\bibinfo {year} {2019})},\ \Eprint {https://arxiv.org/abs/1905.06610} {arXiv:1905.06610 [hep-ph]} \BibitemShut {NoStop}%
\bibitem [{\citenamefont {Ali}\ \emph {et~al.}(2020)\citenamefont {Ali}, \citenamefont {Maiani}, \citenamefont {Parkhomenko},\ and\ \citenamefont {Wang}}]{Ali:2019okl}%
  \BibitemOpen
  \bibfield  {author} {\bibinfo {author} {\bibfnamefont {A.}~\bibnamefont {Ali}}, \bibinfo {author} {\bibfnamefont {L.}~\bibnamefont {Maiani}}, \bibinfo {author} {\bibfnamefont {A.~Y.}\ \bibnamefont {Parkhomenko}},\ and\ \bibinfo {author} {\bibfnamefont {W.}~\bibnamefont {Wang}},\ }\bibfield  {title} {\bibinfo {title} {Interpretation of {{$Y_b(10753)$}} as a tetraquark and its production mechanism},\ }\href {https://doi.org/10.1016/j.physletb.2020.135217} {\bibfield  {journal} {\bibinfo  {journal} {Phys. Lett. B}\ }\textbf {\bibinfo {volume} {802}},\ \bibinfo {pages} {135217} (\bibinfo {year} {2020})},\ \Eprint {https://arxiv.org/abs/1910.07671} {arXiv:1910.07671 [hep-ph]} \BibitemShut {NoStop}%
\bibitem [{\citenamefont {Chen}\ \emph {et~al.}(2025)\citenamefont {Chen}, \citenamefont {Tan}, \citenamefont {Chen}, \citenamefont {Liu},\ and\ \citenamefont {Ping}}]{Chen:2025zbu}%
  \BibitemOpen
  \bibfield  {author} {\bibinfo {author} {\bibfnamefont {X.}~\bibnamefont {Chen}}, \bibinfo {author} {\bibfnamefont {Y.}~\bibnamefont {Tan}}, \bibinfo {author} {\bibfnamefont {Y.}~\bibnamefont {Chen}}, \bibinfo {author} {\bibfnamefont {X.}~\bibnamefont {Liu}},\ and\ \bibinfo {author} {\bibfnamefont {J.}~\bibnamefont {Ping}},\ }\bibfield  {title} {\bibinfo {title} {{$\Upsilon(5S)$ in the unquenched quark model}},\ }\href@noop {} {\  (\bibinfo {year} {2025})},\ \Eprint {https://arxiv.org/abs/2507.13882} {arXiv:2507.13882 [hep-ph]} \BibitemShut {NoStop}%
\bibitem [{\citenamefont {Mehen}\ and\ \citenamefont {Powell}(2013)}]{Mehen:2013mva}%
  \BibitemOpen
  \bibfield  {author} {\bibinfo {author} {\bibfnamefont {T.}~\bibnamefont {Mehen}}\ and\ \bibinfo {author} {\bibfnamefont {J.}~\bibnamefont {Powell}},\ }\bibfield  {title} {\bibinfo {title} {Line shapes in {{$\Upsilon(5S) \to B^{(*)} \bar{{B}}^{(*)}\pi$}} with {{Z}}(10610) and {{Z}}(10650) using effective field theory},\ }\href {https://doi.org/10.1103/PhysRevD.88.034017} {\bibfield  {journal} {\bibinfo  {journal} {Phys. Rev. D}\ }\textbf {\bibinfo {volume} {88}},\ \bibinfo {pages} {034017} (\bibinfo {year} {2013})},\ \Eprint {https://arxiv.org/abs/1306.5459} {arXiv:1306.5459 [hep-ph]} \BibitemShut {NoStop}%
\bibitem [{\citenamefont {Voloshin}(2013)}]{Voloshin:2013ez}%
  \BibitemOpen
  \bibfield  {author} {\bibinfo {author} {\bibfnamefont {M.~B.}\ \bibnamefont {Voloshin}},\ }\bibfield  {title} {\bibinfo {title} {Enhanced mixing of partial waves near threshold for heavy meson pairs and properties of ${{Z}}_b(10610)$ and ${{Z}}_b(10650)$ resonances},\ }\href {https://doi.org/10.1103/PhysRevD.87.074011} {\bibfield  {journal} {\bibinfo  {journal} {Phys. Rev. D}\ }\textbf {\bibinfo {volume} {87}},\ \bibinfo {pages} {074011} (\bibinfo {year} {2013})},\ \Eprint {https://arxiv.org/abs/1301.5068} {arXiv:1301.5068 [hep-ph]} \BibitemShut {NoStop}%
\bibitem [{\citenamefont {Guo}\ \emph {et~al.}(2014)\citenamefont {Guo}, \citenamefont {Mei{\ss}ner},\ and\ \citenamefont {Shen}}]{Guo:2014qra}%
  \BibitemOpen
  \bibfield  {author} {\bibinfo {author} {\bibfnamefont {F.-K.}\ \bibnamefont {Guo}}, \bibinfo {author} {\bibfnamefont {U.-G.}\ \bibnamefont {Mei{\ss}ner}},\ and\ \bibinfo {author} {\bibfnamefont {C.-P.}\ \bibnamefont {Shen}},\ }\bibfield  {title} {\bibinfo {title} {Enhanced breaking of heavy quark spin symmetry},\ }\href {https://doi.org/10.1016/j.physletb.2014.09.043} {\bibfield  {journal} {\bibinfo  {journal} {Phys. Lett. B}\ }\textbf {\bibinfo {volume} {738}},\ \bibinfo {pages} {172} (\bibinfo {year} {2014})},\ \Eprint {https://arxiv.org/abs/1406.6543} {arXiv:1406.6543 [hep-ph]} \BibitemShut {NoStop}%
\bibitem [{\citenamefont {Badalian}\ \emph {et~al.}(2010)\citenamefont {Badalian}, \citenamefont {Bakker},\ and\ \citenamefont {Danilkin}}]{Badalian:2009bu}%
  \BibitemOpen
  \bibfield  {author} {\bibinfo {author} {\bibfnamefont {A.~M.}\ \bibnamefont {Badalian}}, \bibinfo {author} {\bibfnamefont {B.~L.~G.}\ \bibnamefont {Bakker}},\ and\ \bibinfo {author} {\bibfnamefont {I.~V.}\ \bibnamefont {Danilkin}},\ }\bibfield  {title} {\bibinfo {title} {Dielectron widths of the {{S-, D-}}vector bottomonium states},\ }\href {https://doi.org/10.1134/S1063778810010163} {\bibfield  {journal} {\bibinfo  {journal} {Phys. Atom. Nucl.}\ }\textbf {\bibinfo {volume} {73}},\ \bibinfo {pages} {138} (\bibinfo {year} {2010})},\ \Eprint {https://arxiv.org/abs/0903.3643} {arXiv:0903.3643 [hep-ph]} \BibitemShut {NoStop}%
\bibitem [{\citenamefont {Adachi}\ \emph {et~al.}(2023)\citenamefont {Adachi} \emph {et~al.}}]{Belle-II:2022xdi}%
  \BibitemOpen
  \bibfield  {author} {\bibinfo {author} {\bibfnamefont {I.}~\bibnamefont {Adachi}} \emph {et~al.} (\bibinfo {collaboration} {Belle-II}),\ }\bibfield  {title} {\bibinfo {title} {{Observation of $e^+e^-\to\omega\chi_{bJ}(1P)$ and Search for $X_b \to \omega \Upsilon(1S)$ at $\sqrt{s}$ near 10.75 GeV}},\ }\href {https://doi.org/10.1103/PhysRevLett.130.091902} {\bibfield  {journal} {\bibinfo  {journal} {Phys. Rev. Lett.}\ }\textbf {\bibinfo {volume} {130}},\ \bibinfo {pages} {091902} (\bibinfo {year} {2023})},\ \Eprint {https://arxiv.org/abs/2208.13189} {arXiv:2208.13189 [hep-ex]} \BibitemShut {NoStop}%
\bibitem [{\citenamefont {Abumusabh}\ \emph {et~al.}(2025)\citenamefont {Abumusabh} \emph {et~al.}}]{Belle-II:2025iil}%
  \BibitemOpen
  \bibfield  {author} {\bibinfo {author} {\bibfnamefont {M.}~\bibnamefont {Abumusabh}} \emph {et~al.} (\bibinfo {collaboration} {Belle-II}),\ }\bibfield  {title} {\bibinfo {title} {{Search for $e^+ e^- \to \gamma \chi_{bJ}$ $(J = 0, 1, 2)$ near $\sqrt{s} = 10.746$ GeV at Belle II}},\ }\href@noop {} {\  (\bibinfo {year} {2025})},\ \Eprint {https://arxiv.org/abs/2508.16036} {arXiv:2508.16036 [hep-ex]} \BibitemShut {NoStop}%
\bibitem [{\citenamefont {He}\ \emph {et~al.}(2014)\citenamefont {He} \emph {et~al.}}]{Belle:2014sys}%
  \BibitemOpen
  \bibfield  {author} {\bibinfo {author} {\bibfnamefont {X.~H.}\ \bibnamefont {He}} \emph {et~al.} (\bibinfo {collaboration} {Belle}),\ }\bibfield  {title} {\bibinfo {title} {{Observation of $e^+e^- \to \pi^+ \pi^- \pi^0 \chi_{bJ}$ and Search for $X_b \to \omega \Upsilon(1S)$ at $\sqrt{s}=10.867$ GeV}},\ }\href {https://doi.org/10.1103/PhysRevLett.113.142001} {\bibfield  {journal} {\bibinfo  {journal} {Phys. Rev. Lett.}\ }\textbf {\bibinfo {volume} {113}},\ \bibinfo {pages} {142001} (\bibinfo {year} {2014})},\ \Eprint {https://arxiv.org/abs/1408.0504} {arXiv:1408.0504 [hep-ex]} \BibitemShut {NoStop}%
\bibitem [{\citenamefont {Guo}\ \emph {et~al.}(2011)\citenamefont {Guo}, \citenamefont {Hanhart}, \citenamefont {Li}, \citenamefont {Mei{\ss}ner},\ and\ \citenamefont {Zhao}}]{Guo:2010ak}%
  \BibitemOpen
  \bibfield  {author} {\bibinfo {author} {\bibfnamefont {F.-K.}\ \bibnamefont {Guo}}, \bibinfo {author} {\bibfnamefont {C.}~\bibnamefont {Hanhart}}, \bibinfo {author} {\bibfnamefont {G.}~\bibnamefont {Li}}, \bibinfo {author} {\bibfnamefont {U.-G.}\ \bibnamefont {Mei{\ss}ner}},\ and\ \bibinfo {author} {\bibfnamefont {Q.}~\bibnamefont {Zhao}},\ }\bibfield  {title} {\bibinfo {title} {Effect of charmed meson loops on charmonium transitions},\ }\href {https://doi.org/10.1103/PhysRevD.83.034013} {\bibfield  {journal} {\bibinfo  {journal} {Phys. Rev. D}\ }\textbf {\bibinfo {volume} {83}},\ \bibinfo {pages} {034013} (\bibinfo {year} {2011})},\ \Eprint {https://arxiv.org/abs/1008.3632} {arXiv:1008.3632 [hep-ph]} \BibitemShut {NoStop}%
\bibitem [{\citenamefont {Guo}\ and\ \citenamefont {Mei{\ss}ner}(2012)}]{Guo:2012tg}%
  \BibitemOpen
  \bibfield  {author} {\bibinfo {author} {\bibfnamefont {F.-K.}\ \bibnamefont {Guo}}\ and\ \bibinfo {author} {\bibfnamefont {U.-G.}\ \bibnamefont {Mei{\ss}ner}},\ }\bibfield  {title} {\bibinfo {title} {Light quark mass dependence in heavy quarkonium physics},\ }\href {https://doi.org/10.1103/PhysRevLett.109.062001} {\bibfield  {journal} {\bibinfo  {journal} {Phys. Rev. Lett.}\ }\textbf {\bibinfo {volume} {109}},\ \bibinfo {pages} {062001} (\bibinfo {year} {2012})},\ \Eprint {https://arxiv.org/abs/1203.1116} {arXiv:1203.1116 [hep-ph]} \BibitemShut {NoStop}%
\bibitem [{\citenamefont {Guo}\ \emph {et~al.}(2018)\citenamefont {Guo}, \citenamefont {Hanhart}, \citenamefont {Mei{\ss}ner}, \citenamefont {Wang}, \citenamefont {Zhao},\ and\ \citenamefont {Zou}}]{Guo:2017jvc}%
  \BibitemOpen
  \bibfield  {author} {\bibinfo {author} {\bibfnamefont {F.-K.}\ \bibnamefont {Guo}}, \bibinfo {author} {\bibfnamefont {C.}~\bibnamefont {Hanhart}}, \bibinfo {author} {\bibfnamefont {U.-G.}\ \bibnamefont {Mei{\ss}ner}}, \bibinfo {author} {\bibfnamefont {Q.}~\bibnamefont {Wang}}, \bibinfo {author} {\bibfnamefont {Q.}~\bibnamefont {Zhao}},\ and\ \bibinfo {author} {\bibfnamefont {B.-S.}\ \bibnamefont {Zou}},\ }\bibfield  {title} {\bibinfo {title} {Hadronic molecules},\ }\href {https://doi.org/10.1103/RevModPhys.90.015004} {\bibfield  {journal} {\bibinfo  {journal} {Rev. Mod. Phys.}\ }\textbf {\bibinfo {volume} {90}},\ \bibinfo {pages} {015004} (\bibinfo {year} {2018})},\ \Eprint {https://arxiv.org/abs/1705.00141} {arXiv:1705.00141 [hep-ph]} \BibitemShut {NoStop}%
\bibitem [{\citenamefont {Casalbuoni}\ \emph {et~al.}(1993)\citenamefont {Casalbuoni}, \citenamefont {Deandrea}, \citenamefont {Di~Bartolomeo}, \citenamefont {Gatto}, \citenamefont {Feruglio},\ and\ \citenamefont {Nardulli}}]{Casalbuoni:1992yd}%
  \BibitemOpen
  \bibfield  {author} {\bibinfo {author} {\bibfnamefont {R.}~\bibnamefont {Casalbuoni}}, \bibinfo {author} {\bibfnamefont {A.}~\bibnamefont {Deandrea}}, \bibinfo {author} {\bibfnamefont {N.}~\bibnamefont {Di~Bartolomeo}}, \bibinfo {author} {\bibfnamefont {R.}~\bibnamefont {Gatto}}, \bibinfo {author} {\bibfnamefont {F.}~\bibnamefont {Feruglio}},\ and\ \bibinfo {author} {\bibfnamefont {G.}~\bibnamefont {Nardulli}},\ }\bibfield  {title} {\bibinfo {title} {Effective {{Lagrangian}} for quarkonia and light mesons in a soft-exchange-approximation},\ }\href {https://doi.org/10.1016/0370-2693(93)90641-T} {\bibfield  {journal} {\bibinfo  {journal} {Phys. Lett. B}\ }\textbf {\bibinfo {volume} {302}},\ \bibinfo {pages} {95} (\bibinfo {year} {1993})}\BibitemShut {NoStop}%
\bibitem [{\citenamefont {Hu}\ and\ \citenamefont {Mehen}(2006)}]{Hu:2005gf}%
  \BibitemOpen
  \bibfield  {author} {\bibinfo {author} {\bibfnamefont {J.}~\bibnamefont {Hu}}\ and\ \bibinfo {author} {\bibfnamefont {T.}~\bibnamefont {Mehen}},\ }\bibfield  {title} {\bibinfo {title} {Chiral {{Lagrangian}} with {{Heavy Quark-Diquark Symmetry}}},\ }\href {https://doi.org/10.1103/PhysRevD.73.054003} {\bibfield  {journal} {\bibinfo  {journal} {Physical Review D}\ }\textbf {\bibinfo {volume} {73}},\ \bibinfo {pages} {054003} (\bibinfo {year} {2006})},\ \Eprint {https://arxiv.org/abs/hep-ph/0511321} {arXiv:hep-ph/0511321} \BibitemShut {NoStop}%
\bibitem [{\citenamefont {Margaryan}\ and\ \citenamefont {Springer}(2013)}]{margaryanUsingDecayPs41602013}%
  \BibitemOpen
  \bibfield  {author} {\bibinfo {author} {\bibfnamefont {A.}~\bibnamefont {Margaryan}}\ and\ \bibinfo {author} {\bibfnamefont {R.~P.}\ \bibnamefont {Springer}},\ }\bibfield  {title} {\bibinfo {title} {Using the decay {$\psi$}(4160){$\to$} {{X}}(3872) {$\gamma$} to probe the molecular content of the {{X}}(3872)},\ }\href {https://doi.org/10.1103/PhysRevD.88.014017} {\bibfield  {journal} {\bibinfo  {journal} {Physical Review D}\ }\textbf {\bibinfo {volume} {88}},\ \bibinfo {pages} {014017} (\bibinfo {year} {2013})},\ \Eprint {https://arxiv.org/abs/1304.8101} {arXiv:1304.8101 [hep-ph]} \BibitemShut {NoStop}%
\bibitem [{\citenamefont {Eichten}\ \emph {et~al.}(1978)\citenamefont {Eichten}, \citenamefont {Gottfried}, \citenamefont {Kinoshita}, \citenamefont {Lane},\ and\ \citenamefont {Yan}}]{Eichten:1978tg}%
  \BibitemOpen
  \bibfield  {author} {\bibinfo {author} {\bibfnamefont {E.}~\bibnamefont {Eichten}}, \bibinfo {author} {\bibfnamefont {K.}~\bibnamefont {Gottfried}}, \bibinfo {author} {\bibfnamefont {T.}~\bibnamefont {Kinoshita}}, \bibinfo {author} {\bibfnamefont {K.~D.}\ \bibnamefont {Lane}},\ and\ \bibinfo {author} {\bibfnamefont {T.-M.}\ \bibnamefont {Yan}},\ }\bibfield  {title} {\bibinfo {title} {Charmonium: {{The}} model},\ }\href {https://doi.org/10.1103/PhysRevD.17.3090} {\bibfield  {journal} {\bibinfo  {journal} {Phys. Rev. D}\ }\textbf {\bibinfo {volume} {17}},\ \bibinfo {pages} {3090} (\bibinfo {year} {1978})}\BibitemShut {NoStop}%
\bibitem [{\citenamefont {Li}\ and\ \citenamefont {Voloshin}(2013)}]{Li:2013yka}%
  \BibitemOpen
  \bibfield  {author} {\bibinfo {author} {\bibfnamefont {X.}~\bibnamefont {Li}}\ and\ \bibinfo {author} {\bibfnamefont {M.~B.}\ \bibnamefont {Voloshin}},\ }\bibfield  {title} {\bibinfo {title} {Suppression of the {{S-wave}} production of (3/2){$^+$} + (1/2){$^-$} heavy meson pairs in {{e$^+$e$^{-}$}} annihilation},\ }\href {https://doi.org/10.1103/PhysRevD.88.034012} {\bibfield  {journal} {\bibinfo  {journal} {Phys. Rev. D}\ }\textbf {\bibinfo {volume} {88}},\ \bibinfo {pages} {034012} (\bibinfo {year} {2013})},\ \Eprint {https://arxiv.org/abs/1307.1072} {arXiv:1307.1072 [hep-ph]} \BibitemShut {NoStop}%
\bibitem [{\citenamefont {Guo}\ \emph {et~al.}(2009)\citenamefont {Guo}, \citenamefont {Hanhart},\ and\ \citenamefont {Mei{\ss}ner}}]{Guo:2009wr}%
  \BibitemOpen
  \bibfield  {author} {\bibinfo {author} {\bibfnamefont {F.-K.}\ \bibnamefont {Guo}}, \bibinfo {author} {\bibfnamefont {C.}~\bibnamefont {Hanhart}},\ and\ \bibinfo {author} {\bibfnamefont {U.-G.}\ \bibnamefont {Mei{\ss}ner}},\ }\bibfield  {title} {\bibinfo {title} {{Extraction of the light quark mass ratio from the decays $\psi' \to J/\psi \pi^0(\eta)$}},\ }\href {https://doi.org/10.1103/PhysRevLett.103.082003} {\bibfield  {journal} {\bibinfo  {journal} {Phys. Rev. Lett.}\ }\textbf {\bibinfo {volume} {103}},\ \bibinfo {pages} {082003} (\bibinfo {year} {2009})},\ \bibinfo {note} {[Erratum: Phys.Rev.Lett. 104, 109901 (2010)]},\ \Eprint {https://arxiv.org/abs/0907.0521} {arXiv:0907.0521 [hep-ph]} \BibitemShut {NoStop}%
\bibitem [{\citenamefont {Fleming}\ and\ \citenamefont {Mehen}(2008)}]{Fleming:2008yn}%
  \BibitemOpen
  \bibfield  {author} {\bibinfo {author} {\bibfnamefont {S.}~\bibnamefont {Fleming}}\ and\ \bibinfo {author} {\bibfnamefont {T.}~\bibnamefont {Mehen}},\ }\bibfield  {title} {\bibinfo {title} {Hadronic {Decays} of the {$X(3872)$} to {$\chi_{cJ}$} in {Effective Field Theory}},\ }\href {https://doi.org/10.1103/PhysRevD.78.094019} {\bibfield  {journal} {\bibinfo  {journal} {Phys. Rev. D}\ }\textbf {\bibinfo {volume} {78}},\ \bibinfo {pages} {094019} (\bibinfo {year} {2008})},\ \Eprint {https://arxiv.org/abs/0807.2674} {arXiv:0807.2674 [hep-ph]} \BibitemShut {NoStop}%
\bibitem [{\citenamefont {Guo}\ \emph {et~al.}(2013)\citenamefont {Guo}, \citenamefont {Hanhart}, \citenamefont {Mei\ss{}ner}, \citenamefont {Wang},\ and\ \citenamefont {Zhao}}]{Guo:2013zbw}%
  \BibitemOpen
  \bibfield  {author} {\bibinfo {author} {\bibfnamefont {F.-K.}\ \bibnamefont {Guo}}, \bibinfo {author} {\bibfnamefont {C.}~\bibnamefont {Hanhart}}, \bibinfo {author} {\bibfnamefont {U.-G.}\ \bibnamefont {Mei\ss{}ner}}, \bibinfo {author} {\bibfnamefont {Q.}~\bibnamefont {Wang}},\ and\ \bibinfo {author} {\bibfnamefont {Q.}~\bibnamefont {Zhao}},\ }\bibfield  {title} {\bibinfo {title} {{Production of the $X(3872)$ in charmonia radiative decays}},\ }\href {https://doi.org/10.1016/j.physletb.2013.06.053} {\bibfield  {journal} {\bibinfo  {journal} {Phys. Lett. B}\ }\textbf {\bibinfo {volume} {725}},\ \bibinfo {pages} {127} (\bibinfo {year} {2013})},\ \Eprint {https://arxiv.org/abs/1306.3096} {arXiv:1306.3096 [hep-ph]} \BibitemShut {NoStop}%
\bibitem [{\citenamefont {Cheng}(2012)}]{Cheng:2011pb}%
  \BibitemOpen
  \bibfield  {author} {\bibinfo {author} {\bibfnamefont {H.-Y.}\ \bibnamefont {Cheng}},\ }\bibfield  {title} {\bibinfo {title} {Revisiting {{Axial-Vector Meson Mixing}}},\ }\href {https://doi.org/10.1016/j.physletb.2011.12.013} {\bibfield  {journal} {\bibinfo  {journal} {Phys. Lett. B}\ }\textbf {\bibinfo {volume} {707}},\ \bibinfo {pages} {116} (\bibinfo {year} {2012})},\ \Eprint {https://arxiv.org/abs/1110.2249} {arXiv:1110.2249 [hep-ph]} \BibitemShut {NoStop}%
\bibitem [{\citenamefont {Abdesselam}\ \emph {et~al.}(2016)\citenamefont {Abdesselam} \emph {et~al.}}]{Belle:2015tbu}%
  \BibitemOpen
  \bibfield  {author} {\bibinfo {author} {\bibfnamefont {A.}~\bibnamefont {Abdesselam}} \emph {et~al.} (\bibinfo {collaboration} {Belle}),\ }\bibfield  {title} {\bibinfo {title} {{Energy scan of the $e^+e^- \to h_b(nP)\pi^+\pi^-$ $(n=1,2)$ cross sections and evidence for $\Upsilon(11020)$ decays into charged bottomonium-like states}},\ }\href {https://doi.org/10.1103/PhysRevLett.117.142001} {\bibfield  {journal} {\bibinfo  {journal} {Phys. Rev. Lett.}\ }\textbf {\bibinfo {volume} {117}},\ \bibinfo {pages} {142001} (\bibinfo {year} {2016})},\ \Eprint {https://arxiv.org/abs/1508.06562} {arXiv:1508.06562 [hep-ex]} \BibitemShut {NoStop}%
\end{thebibliography}%
\end{document}